\documentclass[jgrga]{agutex2015}

\usepackage{color,soul}
\usepackage{graphicx}
\setkeys{Gin}{draft=false}

\authorrunninghead{ SHEN ET AL.}

\titlerunninghead{LOW-ALTITUDE ION HEATING}

\authoraddr{Corresponding author: Yangyang Shen, Department of Physics and Astronomy, University of Calgary, Calgary, Alberta, Canada. (yangyang.shen@ucalgary.ca)}

\begin{document}


\title{Low-altitude ion heating, downflowing ions, and BBELF waves in the return current region}




\authors{Yangyang Shen\altaffilmark{1},
David J. Knudsen\altaffilmark{1}, 
Johnathan K. Burchill\altaffilmark{1},
Andrew D. Howarth\altaffilmark{1},
Andrew W. Yau\altaffilmark{1},
David M. Miles\altaffilmark{2},
H. Gordon James\altaffilmark{1},
Gareth W. Perry\altaffilmark{1},
and Leroy Cogger\altaffilmark{1}}

\altaffiltext{1}{Department of Physics and Astronomy, University of Calgary, Calgary, Alberta, Canada.}
\altaffiltext{2}{Department of Physics and Astronomy, University of Iowa, Iowa City, Iowa, USA}


\begin{abstract}

Heavy (O$^{+}$) ion energization and field-aligned motion in and near the ionosphere are still not well understood. Based on observations from the CASSIOPE Enhanced Polar Outflow Probe (e-POP) at altitudes between 325 km and 730 km over one year, we present a statistical study (24 events) of ion heating and its relation to field-aligned ion bulk flow velocity, low-frequency waves and field-aligned currents (FACs). The ion temperature and field-aligned bulk flow velocity are derived from 2-D ion velocity distribution functions measured by the suprathermal electron imager (SEI) instrument. Consistent ion heating and flow velocity characteristics are observed from both the SEI and the rapid-scanning ion mass spectrometer (IRM) instruments. We find that transverse O$^{+}$ ion heating in the ionosphere can be intense (up to 4.5 eV), confined to very narrow regions ($\sim$ 2 km across B), is more likely to occur in the downward current region, and is associated with broadband extremely low frequency (BBELF) waves. These waves are interpreted as linearly polarized perpendicular to the magnetic field. The amount of ion heating cannot be explained by frictional heating, and the correlation of ion heating with BBELF waves suggest that significant wave-ion heating is occurring and even dominating at altitudes as low as 350 km, a boundary that is lower than previously reported. Surprisingly, the majority of these heating events (17 out 24) are associated with core ion downflows rather than upflows. This may be explained by a downward-pointing electric field in the low-altitude return current region.

\end{abstract}


\begin{article}

\section{Introduction}

 It has been several decades since the discovery of ionospheric O$^{+}$ ions flowing outward into the magnetosphere \citep{shelley1972}. These heavy ions play an important role in diverse magnetospheric regions and exhibit a range of energies from several eV to tens of keV \citep{chappell1988}. One of the most intriguing and fascinating issues concerning the ion outflow is how the O$^{+}$ ions are energized enough to escape Earth's gravity. There exists an extensive body of research documenting such ion energization processes \citep[see reviews by][and references therein]{klumpar1986,yau1997,andre1997,moore1999,moore2010,strangeway2005}. Ion acceleration perpendicular to the magnetic field is commonly observed in these studies. These transversely accelerated ions (TAI) are observed by satellites at high altitudes as ``conics", which refers to a cone-shaped ion distribution in velocity space with maximum flux intensities oblique to the magnetic field line. Conics result from purely transverse acceleration to the local magnetic field at low altitudes with subsequent upward mirror acceleration parallel to $\vec{B}$ due to the diverging geomagnetic field \citep{sharp1977}. Based on the early observations at high altitudes \citep{klumpar1979} and an application of the first adiabatic invariant, source altitudes well below 1000 km are inferred. When the ions are far above the heating source region, or further accelerated by parallel electric fields, conical distributions give way to beam-like distributions.
 
 At relatively low altitudes (\textless 2000 km), these TAIs take more complicated forms as observed by many satellites, sounding rockets and ground-based radars. Ions can resonantly interact with waves to produce bulk heating or tail heating. In the tail heating scenario, when the ion velocity matches the wave phase velocity, the fraction of the ion population in the tail near that velocity is resonantly accelerated by waves \citep{tsurutani1997}. In the bulk heating case, the whole ion population interacts with waves through cyclotron resonance \citep{lynch1999}. The early sounding rocket measurements from \citet{whalen1978} and \citep{yau1983} revealed enhanced ion fluxes in the high-energy tail with energies up to 500 eV and pitch angles near 90\deg~at 500 km altitude. This tail heating feature was also observed by TOPAZ 3 and AMICIST sounding rockets between 500 and 1100 km and was found to be correlated with lower hybrid solitary structures (LHSS) within density-depleted cavities having diameters of tens of meters perpendicular to $\vec{B}$ \citep{arnoldy1992,kintner1992,lynch1996,lynch1999}. 
 
 On the other hand, the Freja satellite observed core ion (\textless 20 eV) bulk heating perpendicular to the magnetic field by broadband extremely low frequency (BBELF) electric fields and without any apparent tail heating feature at 1700 km altitude \citep{norqvist1996,knudsen1998}. These bulk characteristics were also determined from ground-based radar measurements of anisotropic ion temperatures (perpendicular-to-B ion temperature in excess of field-aligned temperature) associated with strong electric fields in the topside ionosphere \citep{wahlund1992}. A recent study by \citet{archer2015} demonstrated that the ratio of perpendicular to parallel ion temperatures can reach as high as 5 at 500 km altitude as observed by the Swarm satellites. The anisotropic value is far beyond the theoretical prediction from DC electric field-driven frictional heating \citep{loranc1994}. Many studies have shown that ion bulk heating and tail heating can occur both simultaneously and independently \citep{garbe1992,moore1996a,moore1996b,andre1998,lynch2007}. Based on GEODESIC sounding rocket observations between 700 and 1000 km altitudes, \citet{burchill2004} proposed that the estimated 5-10 eV ion temperature enhancements of TAIs within lower-hybrid cavities are actually due to bulk heating of the dominant (90\%) cold Maxwellian ions. 
 
 Given the complexity of the observed ion energization features, it is not inconceivable that a number of complex processes are at work simultaneously or in sequence. The relative importance of these processes depends upon conditions within different regions and at different altitudes, as well as on geomagnetic activity levels, and has still not been fully resolved, particularly in the topside ionosphere \citep{moore1999}. It is known that the heavy O$^{+}$ ions undergo stepwise energization processes starting from below the F region \citep{strangeway2005,fernandes2016}. At high altitudes (\textgreater 1000 km) wave-ion heating perpendicular to the magnetic field and the subsequent parallel acceleration due to the magnetic mirror force and field-aligned electric fields accelerate ionospheric ions to escape speeds and expel them into the magnetosphere \citep{andre1997,moore2010}. 
 
 Below 1000 km, two mechanisms are often cited to help seed enough thermal ionospheric ions into the wave-ion heating region \citep{wahlund1992,strangeway2005}. The first is frictional heating from relative motion between convection ion drifts and neutrals, increasing perpendicular ion temperatures and the plasma scale height, which results in thermal expansion of ions in the topside ionosphere \citep{loranc1994,wilson1994}. The second is associated with soft electron precipitation, which directly energizes the thermal ionospheric electrons instead of the thermal ions since the Coulomb collision cross section is inversely proportional to the mass and relative speed between collisional particles \citep{schunk2009}. The resulting thermal expansion of electrons and the density gradient create an upward-pointing ambipolar electric field that can accelerate the heavy ions upward along the magnetic field line \citep{whitteker1977,liu1995,seo1997}. Based on FAST satellite observations, \citet{strangeway2005} found high correlations of ion upflow with both soft electron precipitation and Poynting flux. Those authors suggest these two mechanisms as the ultimate drivers of ion energization processes, since Poynting flux dissipates as frictional heating at low altitudes and both of them can seed low-frequency waves through instabilities at high altitudes\citep{strangeway2005}.
  
 Wave-ion heating is not considered as important as frictional heating below 1000 km, partly due to the collisional nature of the ionosphere which can damp waves, and partly due to lack of simultaneous and appropriate wave and plasma instruments operating at very low altitudes, especially below 500 km. Using combined EISCAT UHF and VHF radar observations of the topside auroral ionosphere, \citet{wahlund1992} concluded that ion upflows during periods of high convection electric fields, known as type 1, are caused by strong frictional heating of the ions. Based on DE-2 satellite ion drift meter measurements, \citet{loranc1991} statistically investigated vertical ion bulk flow velocities between 200 and 1000 km. They suggested that the upward ion drifts were probably driven by frictional heating below 400 km. Analyzing similar ion drift meter data onboard the HILAT satellite, \citet{tsunoda1989} suggested that the upward ion drifts co-located in regions of the convective velocity shear were most likely transversely accelerated ions or ion conics. Due to the small convective velocity, they suggested that the heated ions cannot be explained by frictional heating alone but require additional sources such as plasma waves. 
  
  With dedicated low-energy ion and wave measurements, on the other hand, numerous sounding rocket observations indicate that wave-ion heating contributes significantly to the ion energization process in the topside ionosphere below 1000 km. Correlation between the tail or core ion heating and the BBELF waves or the lower hybrid waves has been reported between 325 km and 1000 km by \citet{whalen1978} and \citet{yau1983}, and by subsequent sounding rocket missions such as MARIE \citep{kintner1986}, TOPAZ series \citep{moore1986,garbe1992,arnoldy1992,kintner1992}, AMICIST \citep{lynch1996}, ARCS-4 \citep{moore1996b}, GEODESIC \citep{burchill2004}, SIERRA \citep{lynch2007}, SERSIO \citep{frederick2007},VISIONS \citep{collier2015}, and MICA \citep{fernandes2016}. Despite relatively weak (0.2 eV) ion temperature increases, \citet{fernandes2016} suggests that ion heating by plasma waves may occur at 325 km altitude based on observations from the MICA sounding rocket. Prior to this the lowest reported altitude of wave-particle heating was 450 km \citep{whalen1978}. While limited in terms of sample number, these examples support the idea that wave-ion heating plays an important role at low altitudes. In this study, we will show that very strong wave-ion heating (up to an ion temperature of 4.5 eV) clearly occurs at altitudes as low as 350 km. 
  
  While it is apparent that wave electric fields heat the ions perpendicularly to the magnetic field, the wave modes responsible for the transverse heating remain uncertain. BBELF waves and lower-hybrid waves are the most common candidates \citep{arnoldy1992,kintner1992,kintner1996,lynch1996,lynch2007,andre1994,andre1998,moore1999}. Based on 20 months of observations from the Freja satellite at altitudes of up to 1700 km, \citet{norqvist1998} found that nearly 90\% of all O$^{+}$ ion heating events and near 95\% of the total O$^{+}$ upflow are caused by ion energizations associated with the BBELF waves and that post-midnight and morning sectors are the preferential regions for BBELF wave-ion resonance heating. \citet{knudsen1998} found that these ion heating events are correlated with soft electron precipitation as well. A similar conclusion was made by \citet{lynch1996}, who used the nightside sounding rocket measurements from AMICIST to illustrate that ion heating events associated with BBELF are much stronger than heating events within the LHSS. \citet{lynch2002} found from FAST satellite measurements that ion energization from BBELF waves is frequently observed within downward current regions with downward-pointing electric fields from the FAST satellite measurements. BBELF waves may comprise different wave modes such as Alfven waves below the ion cyclotron frequency, ion acoustic waves \citep{wahlund1998}, electrostatic waves near the ion cyclotron frequency \citep{bonnell1996} and electromagnetic waves near the ion cyclotron frequency \citep{andre1997,moore1999}. These waves may be seeded from current-driven instability \citep{kindel1971} or shear-driven instability \citep{ganguli1994}. In general, left-hand circularly or linearly polarized waves (which can be decomposed into both left and right circularly polarized waves) near the ion cyclotron resonance frequencies are important for the transverse ion heating process. 
  
  Field-aligned ion bulk flow velocities in the topside ionosphere show different characteristics as seen by ground-based radars versus in-situ sounding rocket measurements. Based on 7 years of EISCAT VHF radar observations of upflowing and downflowing ions, \citet{endo2000} suggested that ion upflows often become discernible only above 600 km altitude for the radar measurements. At lower altitudes, the averaged ion upflow velocity is below the level of detectability at the radar's sampling resolution (approximately 1-min temporal and tens of km spatial resolution). Statistical results from the EISCAT Svalbard Radar (ESR) reported by \citet{buchert2004}, on the other hand, demonstrated that the flow is often upward above the F-region density peak and downward below the peak. This diverging flow is interpreted as being related to transport processes resulting from thermal electron heating and ionization due to soft electron precipitation. Under the effects of gravity and the pressure gradient, ions are driven downward below the F-region density peak.
  
  However, observations from several sounding rockets clearly showed that downflowing ions can occur above the F-region peak, sometimes with but usually without concurrent ion upflows. \citet{kintner1986} reported strong perpendicular ion energization (up to 300 eV) associated with plasma waves between 500 and 900 km altitudes using measurements from the MARIE auroral sounding rocket. Ion energy enhancements were obvious both in the downward-going ions (0-90\deg~pitch angle) and perpendicular ions (90\deg-120\deg~pitch angle). Near 600 km altitude, the SIERRA nightside auroral sounding rocket observed low-energy (10 eV) upflowing ions (100\deg-140\deg~pitch angle) up to 2 km/s with a core ion temperature of 1 eV at the poleward edge of an auroral arc \citep{lynch2007}. Relatively high-energy (10-100 eV) ions were observed going downward with a tail temperature of tens of eV. These ions were interpreted as the reflection from high-altitude downward electric fields for upgoing transversely heated ion conics seeded by the low altitude upflow. 
  
  A similar pair of counter-streaming ion populations was observed by \citet{garbe1992} using TOPAZ-2 sounding rocket measurements near 700 km altitude. They reported that the ambient ions (mainly O$^{+}$), which were heated up to 3 eV, were streaming down the magnetic field towards the atmosphere with speeds between 3-5 km/s. These ions sometimes were accompanied by an upgoing suprathermal tail (mainly H+) with temperatures of tens of eV. \citet{moore1996b} analyzed the ion distribution function measurements from the ARCS-4 sounding rocket at altitudes between 500-550 km and revealed that core ion heating close to 0.6 eV in temperature was coincident with nearly 1 km/s ion downflows as well as 2-3 km/s horizontal flows. \citet{collier2015} reported remotely imaged energetic neutral atoms (ENAs) in the polar cusp with energies of near 100 eV using coordinated rocket (VISIONS) and ground-based camera observations. The origin of these upgoing ENAs was identified as downward and transversely-moving ions of energies up to 100 eV. This transverse ion heating and downflowing ions were explained by the effects of downward-pointing electric fields at high altitude.

Understanding the ion energization processes near the topside ionosphere proves to be critical for our knowledge of the field-aligned ion upflow process. Whereas there are statistical studies of wave-ion heating from high-altitude satellites, no statistical study exists at very low altitudes with dedicated low-frequency wave and low-energy plasma instruments. In addition, the majority of the aforementioned studies suffer to some extent from the coarse temporal resolution (at best 1 s) which leads to poorly-resolved structures of ion temperature and flow within the ion heating region. In this study, we will present a statistical investigation of high-resolution (up to 100 samples per second (sps)) low-energy (\textless 100 eV) ion distribution function measurements as well as simultaneous low-frequency (up to 30 kHz) wave electric field observations from the CASSIOPE Enhanced Polar Outflow Probe (e-POP) satellite at altitudes of 325 -730 km. We will show that the observed strong heating events (up to 4.5 eV) are related to the BBELF waves and cannot be explained by frictional heating, suggesting significant wave-ion heating at these altitudes. Contrary to what would be expected from mirror-force acceleration of the heated ions, the majority of these heating events are associated with core ion downflow rather than upflow. The Instrumentation section gives a brief description of the data set used in this study. The Data section presents two events in detail and the statistical results in general. Finally the Discussion and Conclusion sections expand on the observations and interpretations, and summarize our findings.


\section{Instrumentation and Data}

In this study, we investigate low-energy ($\textless$ 100 eV) ion, magnetic field, wave electric field and auroral optical data measured from the e-POP satellite between March 2015 and March 2016. e-POP is an 8-instrument scientific payload on the multi-purpose CASSIOPE satellite \citep{yauandjames2015}, launched on 29 September 2013 into a polar elliptical orbit plane with an inclination of 81\deg, a perigee of 325 km and an apogee of 1500 km. The satellite is three-axis stabilized with an orbital velocity of 7-8 km/s. The primary objective of the e-POP mission is to study polar outflow and the associated microscale ion energization processes. 

\begin{figure}
\includegraphics[scale=0.5,width=9cm]{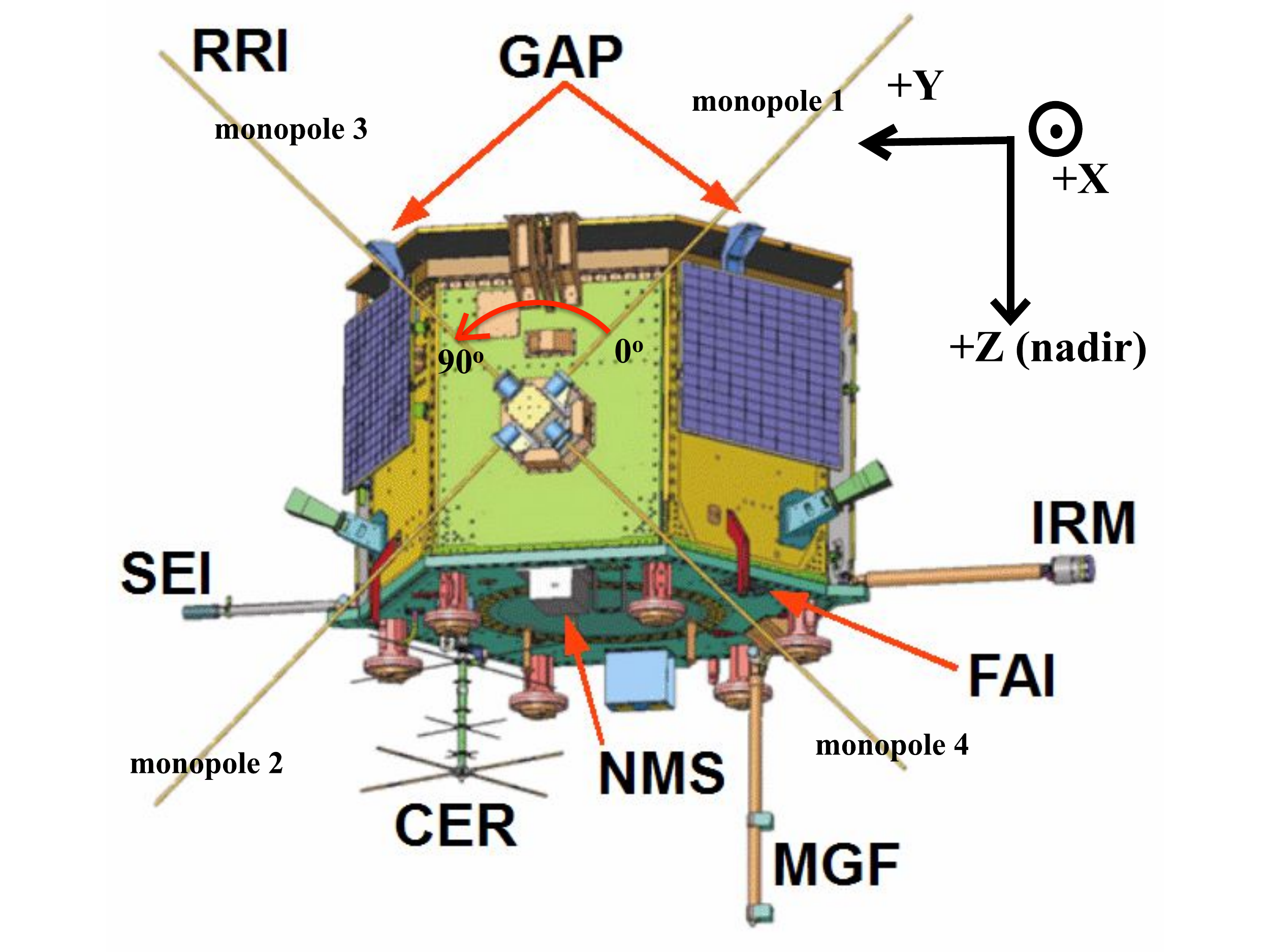}
\caption{Placement and orientation of instruments onboard e-POP in nadir-pointing attitude, where the satellite +X points towards the spacecraft ram direction, which is out of the paper, +Z points towards nadir and +Y completes the right-handed system and aligns with the boom that holds the SEI instrument. The monopole positions of RRI along with the polar coordinates defined within the dipole plane are also shown (red arrow with polar angles). This figure is adapted and modified from figure 2 in \citet{yauandjames2015}. }
\label{epop}
\end{figure}

e-POP carries in total eight scientific instruments, five of which are used in this study: the suprathermal electron imager (SEI) \citep{knudsen2015}, the rapid-scanning ion mass spectrometer (IRM) \citep{yau2015}, the magnetic field instrument (MGF) \citep{wallis2014}, the radio receiver instrument (RRI) \citep{james2015}, and the fast auroral imager (FAI) \citep{cogger2014}. Figure~\ref{epop} depicts the placement and orientation of different sensors on the spacecraft. In the spacecraft coordinate system when in nadir-pointing attitude, +X points towards the spacecraft ram direction, +Z points towards the nadir, and +Y completes the right-handed system and is aligned with the boom that holds SEI. All e-POP data used in this study were taken while in nadir-pointing attitude. Data sampling for the high-data-rate instruments such as SEI, IRM and RRI are often limited to 20 minutes per day, in order to maximize the data sampling resolution despite the limited daily telemetry bandwidth. Such limitation is due to the small number of available daily data downlink orbit passes and the limited availability of the high-speed $K_a$-band downlink because of technical issues. Not every instrument operates each day. 

High-resolution (up to 100 sps) measurements of ion upflow velocities and temperatures are obtained from the e-POP SEI instrument, whose primary purpose is the measurement of suprathermal electrons but it can be configured to measure positive ions. Note that SEI ion mode and electron mode are mutually exclusive. Details on ion measurements from the SEI can be found in \citet{shen2016}. SEI comprises a hemispherical electrostatic analyzer (HEA), a micro-channel plate (MCP), and a phosphor screen coupled to a charge-coupled device (CCD) by fibre optic components. The focusing system of HEA allows SEI to image in both energy and arrival angle with a high frame rate, in contrast to the commonly used top-hat analyzer \citep{carlson1982}, which must step through energy with time. In addition, for the same given hemispherical potential, the SEI can measure particle energies much lower than what the top-hat allows \citep{knudsen2015}. The sensor is intentionally biased to a negative potential (usually near -5 V) with respect to the spacecraft to compensate for any possible positive charging of the spacecraft and to draw in low-energy ions. By controlling the radial electric field between hemispherical electrodes, it measures two-dimensional ion velocity distribution functions (in the spacecraft X-Z plane), from which the bulk kinetic energy (taking into account voltage settings, in general \textless100 eV for this study) and arrival angles (field of view spanning 360\deg) of incoming particles can be derived. Distribution images are 64 pixels in diameter at higher image resolution (Hi-res mode) or 32 pixels in diameter at lower resolution (Normal mode). Both types of images are used in this study. 

At high latitudes where the satellite ram direction is nearly perpendicular to the local magnetic field direction, ion flow velocity parallel to the geomagnetic field manifests as an offset of the first moment of the ion distribution from the nominal anti-ram direction. The along-track component (usually in north-south direction) of the plasma convection velocity is usually assumed to be much smaller than the satellite velocity (7-8 km/s), which is reasonable when the horizontal ion velocity vector is mainly in the east-west direction in the east-west elongated aurora arcs. The nominally vertical component of the ion velocity $V_{c}$ is approximated as: 
\begin{equation}
\label{eq:vcross}
V_{c}  = (V_{s} + V_{p}) \times \tan{\theta} \simeq V_{s} \times \tan{\theta}
\end{equation}
where all the quantities are calculated in the SEI image frame. In this calculation we neglect $V_{p}$, which is the plasma convection velocity in the anti-ram direction, as we assume it is much smaller than the satellite ram velocity $V_{s}$. $\theta$ is the angle between the direction aligning the centroid position with the image center and the anti-ram direction. The ion velocity parallel to $\vec{B}$ can be obtained by projecting $V_{c}$ onto the magnetic field direction. The angular resolution of the centroid variation is of the order of 0.2\deg, corresponding to a vertical velocity resolution of $\sim$25 m/s. The uncertainty in the vertical ion velocity, mainly owing to neglect of $V_{p}$ in Equation~\ref{eq:vcross}, is in general less than 60 m/s for upflow velocities when the deflection angle is less than 10\deg~\citep{shen2016}. However, the uncertainty reaches 800 m/s in an extreme case when the deflection angle reaches 22\deg, corresponding to a vertical ion velocity of 3.2 km/s, if we assume an along-track ion convection velocity of 2 km/s.

We also estimate ion temperatures from SEI by determining the distribution function width. Through a Monte-Carlo charged particle ray tracing simulation \citep{burchill2010} adapted to the SEI instrument, we find linear relations between the distribution's full width at half maximum (FWHM, measured in pixels) and ion temperature (eV) at different deflecting voltages: $T_{i}$/FWHM = 0.51 and 0.84 in the case of $V_{inner}$ = -98 V and  -502 V, respectively, in Hi-res mode, where $V_{inner}$ is the deflection voltage applied across the hemispherical domes. The ratio of $T_{i}$/FWHM in Normal mode is often double that value in Hi-res mode under the same deflection voltage. The estimated ion temperature is cross-checked through forward modeling of flight images and comparing slopes of the simulated and observed distributions. It is noted that the measured flight image is the convolution of the real ion distribution with the instrumental systematic response. Therefore factors other than the ion temperature such as the spacecraft potential can result in FWHM changes in the measurements, especially under relatively small deflection voltage when we are measuring several-eV ions. However, this effect is insignificant for the high-voltage ($V_{inner}$$\ge$ 100 V) setting when the majority of the measured ions have energies of tens of eV or larger. The effect of spacecraft potential on the perpendicular FWHM measurements is shown in Appendix A.

In this study we use measurements from the e-POP IRM as an independent check on ion velocity and temperature measurements from the SEI, though the IRM instrument is routinely used to measure the ionospheric ion composition and ion energies. The IRM instrument samples the ion velocity distribution with 100 frame-per-second time resolution in the energy-per-charge range of $\sim$1-90 eV/q using a HEA similar to SEI but with discrete anode pixel detector instead of CCD detector \citep{yau2015,yau2016}. It has a time-of-flight (TOF) gate to resolve the ion composition in the range of 1 to 40 atomic mass units per charge (AMU/q). As compared to 64$\times$64 pixels (Hi-res mode) in SEI, the ion detector in IRM contains in total 64 discrete pixels, arranged in 8 angular pixel sectors of 8 energy pixels each. Instead of applying a fixed deflecting voltage in the HEA during each orbit as in the case of SEI, the IRM sensor sweeps several voltages across the HEA domes from $\sim20$ V to $\sim350$ V during each measurement. The peak ion energy-per-charge varies approximately as the square of the pixel radius for a given voltage. The angular sectors allow one to determine arrival angle within the plane of the entrance aperture. 

We use a similar algorithm as described above for the SEI to derive the ion field-aligned bulk flow velocity from the IRM ion distribution function measurements. The quantity $\theta$, which is the angular deflection of the centroid position (with respect to the image center) away from the anti-ram direction, is calculated from the first angular moment of the ion distribution. The baseline centroid angle was subtracted during this calculation. The average perpendicular bulk flow velocity and temperature are inferred from the first and second perpendicular-to-B moments of the distribution. The comparison with SEI is made using instrument-frame quantities and does not require that IRM measurements be calibrated in terms of physical units. We only use pixels that have significant signal for the calculation. For all the events we report here, IRM observations reveal that approximately 90-99\% of ions are O$^{+}$ (mixed with a small amount of N$^{+}$ and NO$^{+}$ ions), and about 1-10\% of ions are H$^{+}$ and He$^{+}$. Lighter ions (H$^{+}$, He$^{+}$) are excluded in order to restrict our calculation mainly to O$^{+}$ ions. We will show below only the IRM ion measurements under the highest voltage setting ($V_{deflection}$ = 350 V), which includes ions with energies from sub-eV up to about 75 eV. Lower-voltage setting ion measurements are also available and are in general consistent with the high-voltage results. 

The RRI comprises a four-channel digital receiver fed by four 3-meter monopoles, which can be configured to function as a crossed-dipole antenna lying in the plane parallel to the spacecraft Y-Z plane, in addition to an electronics unit. In VLF mode, it measures wave electric fields nominally from 10 Hz up to near 30 kHz, with a sampling rate of 62.5 kHz. However, it can also be configured to extend the lower end of the frequency range to below 10 Hz. The in-phase and quadrature components in each dipole voltage enable us to derive the wave power spectral density (PSD) and polarization information in the RRI dipole plane after conducting complex Fast Fourier Transform (C-FFT) analysis of the electric field time series. A window length of 0.26 second is chosen for each C-FFT calculation after weighting the time series by a Hanning window. The resulting frequency resolution in the time-frequency spectrogram is approximately 3.8 Hz. 

The polarization ellipse of the wave electric fields can be expressed in terms of the orientation angle ($\psi$) and the ellipticity angle ($\chi$) \citep{collett2005}, which are defined by the following equations: 
\begin{equation}
\label{eq:orient}
\tan{2\psi} = \frac{2 E_{0x} E_{0y}} {{E_{0x}}^2 -{E_{0y}}^2} \cos{\delta}, \qquad 0 \leq \psi \leq \pi 
\end{equation}
\begin{equation}
\label{eq:ellipse}
\tan{2\chi} = \frac{2 E_{0x} E_{0y}} {{E_{0x}}^2 +{E_{0y}}^2}  \sin{\delta}, \qquad -\pi/4 \leq \chi \leq \pi/4
\end{equation}
 where $E_{0x}$ and $E_{0y}$ are the amplitudes of the orthogonal wave electric fields and $\delta$ is the phase difference between them after conducting C-FFT analysis. The voltage values are used to represent the amplitudes of wave electric fields. Wave voltage amplitudes can be converted to wave electric field amplitudes by dividing by the dipole effective length. For the present case of a dipole that is short compared to the assumed electromagnetic wavelength, the effective length of our 6-m dipoles is 3 m. The ellipticity angle and orientation angle spectrograms that we show are averaged to a frequency resolution of 7.6 Hz and a time resolution of 0.52 s. 
 
 The waves are nominally linearly polarized if the ellipticity angle $\chi$ approaches 0\deg~and are circularly polarized if approaching $\pm$45\deg. Whether a given ellipticity angle $\chi$ is left-hand or right-hand polarized with respect to the magnetic field direction depends on the relative angle between the local magnetic field and the dipole plane normal direction, which is along the satellite ram direction. The magnetic field is usually in the zenith-pointing direction in our nightside auroral events and is at a small angle to the crossed-dipole plane. In the polar coordinates defined within the cross-dipole plane (Figure~\ref{epop}), the angle between the magnetic field direction and the cross-dipole is roughly 45\deg. The major axis of the 2-D polarization ellipse is parallel to the magnetic field if the orientation angle $\psi$ approaches 45\deg~and is perpendicular to the magnetic field if approaching -45\deg. It is emphasized that we can only measure the projection of 3-D wave electric fields onto the 2-D dipole plane. Therefore the polarizations derived from above cannot fully resolve the complexity of the plasma wave total electric field vector.

The MGF provides vector magnetic field measurements 160 times per second to a resolution of 0.625 nT in the spacecraft frame, from which we can derive the magnetic perturbations after subtracting the International Geomagnetic Reference Field (IGRF) model. Based on Ampere's law, the field-aligned current (FAC) directions can be inferred from the gradients (positive or negative) of the magnetic field perturbation in the cross-track component (spacecraft Y component) together with the spacecraft velocity direction (northward or southward), nominally. In this calculation, we assume a sheetlike current structure and a spacecraft trajectory that is moving through this structure at an oblique angle. The calculation also assumes that the current sheet is static over both the spatial scale and time scale required to collect the date used in the calculation.

The FAI consists of two charge-coupled device (CCD) cameras with one measuring the 630 nm emissions of atomic oxygen in aurora and the other observing prompt auroral emissions in the 650 to 1100 nm range. The nadir-pointing near-infrared camera which provides one image per second with a spatial resolution of a few km is used in this study. The FAI images are presented in the form of dynamic auroral movies in the Supplementary Information (SI) with a frame rate of one per second. We trace the magnetic footprint of e-POP using the IGRF model from the satellite altitude down to 110 km and mark the footprint on the FAI images to show the auroral context of ion heating events. 

We conduct a statistical survey based on one year's worth of e-POP data to identify strong ion heating signatures (ion temperature potentially larger than 0.5 eV) identified using the SEI instrument. There are in total 134 orbits of 5-10 minutes duration each with SEI data of sufficient quality for analysis; 17 of them show in total 24 clear O$^{+}$ ion heating events. There are 18 ion heating events that have simultaneous low-frequency wave observations from the RRI instrument, exhibiting BBELF wave emissions in all cases. The majority of them have simultaneous observations from the IRM, the MGF and the FAI to provide complementary perspectives on the ion heating events. In the following section, we first present detailed observations of two examples, in the nightside auroral region and in the dayside cleft region, respectively. The statistical results are also given in this section.


\section{Observations}
\subsection{Example in the Nightside Aurora}

The first ion heating example occurred near magnetic midnight when the e-POP payload passed through multiple regions of auroral emissions in the southern hemisphere on May 18, 2015. The satellite travelled from Altitude Adjusted Corrected Geomagnetic (AACGM) latitude -69\deg~to -76\deg~between 22:29:13 and 22:31:11 UT at altitudes from 427 km to 397 km. Near 22:30:09 UT, the e-POP SEI observed strong ion heating signatures at an altitude of 412 km. SEI was operating in Hi-res mode and had a deflection voltage of $V_{inner}$ = -502 V, corresponding to ion energies of up to 230 eV. In-flight ion images show that most ions had energies below 100 eV. SEI data were integrated to 0.1 second resolution in order to reduce noise. Complementary observations from the MGF, IRM, FAI and RRI instruments were simultaneously accessible during this time. The Kp index was 5+ during this event. O$^{+}$ ions were the major ion species according to the IRM.

\begin{figure}
\includegraphics[width=8cm,scale=0.5]{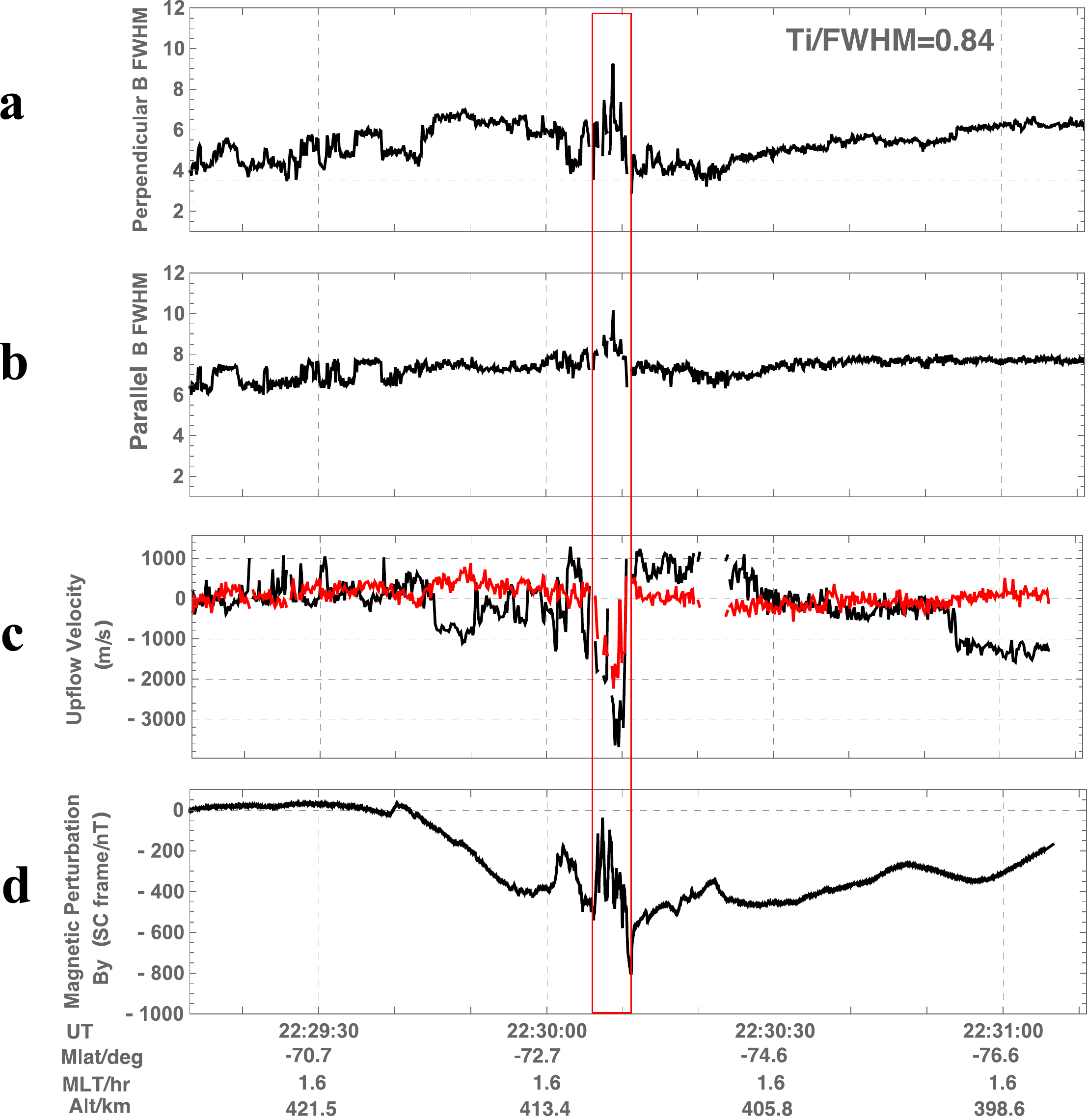}
\caption{e-POP SEI (ion mode) and MGF observations of a core ion heating and downflow region on 18 May 2015 at an altitude of 410 km. a) Full width at half maximum (FWHM, in pixels) of the core ion distribution in the direction perpendicular to $\vec{B}$. The sudden increase in FWHM near 22:30:09 UT represents nearly a 4.3 eV increase in the perpendicular ion temperature, which lasts for 0.2 seconds, corresponding to less than 2 km along the satellite track. b) FWHM fluctuations parallel to $\vec{B}$. The enhancement near 22:30:09 UT represents a parallel ion temperature increase of less than 3 eV. c) Field-aligned ion bulk flow velocities (averaged to 10 frames per second) measured from both the SEI (black) and IRM (red) instruments. Both SEI and IRM observe downward ion bulk flows of over 2 km/s. d) Magnetic field perturbations in the horizontal cross-track direction, which is approximately eastward for positive values. Positive gradients of $\Delta B_y$ along the satellite track represent downward currents in this case. In-between the apparent large-scale upward and downward current sheets, highly-fluctuating magnetic perturbations up to 800 nT are observed within the ion heating region (red box region).}
\label{firstsummary}
\end{figure}

 Figure~\ref{firstsummary} summarizes the observations from the SEI and MGF instruments, which characterize ion temperatures, ion field-aligned flow velocities, and magnetic perturbations in the spacecraft +Y (nominally east) component. Figure~\ref{firstsummary}a shows the perpendicular (to $\vec{B}$) FWHM variations (in pixels) of the measured ion distributions as a function of time. As mentioned in the previous section, ion temperatures from FWHM are calibrated using a Monte-Carlo simulation of the instrument. Linear relations between FWHM variations and ion temperatures are obtained for different instrumental settings. In the case of $V_{inner}$ = -502 V in Hi-res mode, we use the conversion value of 0.84 as an estimate of the upper limit of increase in ion temperature for each pixel increase in FWHM. The sudden increase in FWHM near 22:30:09 UT represents nearly a 4.3 eV increase in the perpendicular ion temperature, which lasts only 0.2 seconds, corresponding to less 2 km along the satellite track. Note that this 2-km wide ion heating region resides within a wider, moderately-heated region spanning approximately 40 km, indicated by the rectangular red box region. 
 
 Figure~\ref{firstsummary}b presents the parallel (to $\vec{B}$) FWHM fluctuations during the measured period. The enhancement in parallel FWHM near 22:30:09 UT represents a parallel ion temperature increase of less than 3 eV. However, we are cautious in interpreting the parallel ion temperatures. Monte-Carlo simulations indicate that the parallel FWHM variations may be affected by other factors, such as horizontal ion drifts and spacecraft potential. Both parallel and perpendicular ion temperatures during the heating time are further cross-checked through forward modeling, by matching the simulated ion distribution function slope with observations, as will be described in the following part of this section. Figure~\ref{firstsummary}c displays the field-aligned ion bulk flow velocities measured from both the SEI (black line) and IRM (red line) instruments. Ion flow velocities derived from the IRM instrument are also averaged to 10 frames per second (fps) from near 60 fps. Inside the ion heating region, both SEI and IRM observe downward ion bulk flows of over 2 km/s. The maximum downflowing ion velocity observed by the SEI is over 3 km/s, which might be overestimated by several hundred m/s as described in the previous section, considering potentially large along-track ion drift velocities indicated by the IRM and FAI data in the following part of this section. In addition, the SEI shows ion upflow velocities of up to 1 km/s on both sides of the downward flow region. This feature is also visible in the IRM-derived velocities, though showing less structures and smaller amplitudes. Discrepancies between SEI-derived and IRM-derived ion drift velocities are potentially due to compromised counting statistics and angular resolution in the ion measurements from IRM in order to allow ion mass resolution. This limits the uncertainty and resolution of IRM-derived ion flow velocities to be no finer than at least 500 m/s. Velocity variations estimated from IRM using Equation~\ref{eq:vcross} cannot be interpreted as of physical origin if they are smaller than 500 m/s.
 
 Figure~\ref{firstsummary}d presents the magnetic field perturbations in the horizontal cross-track direction, which is approximately eastward for positive values. Positive gradients of $\Delta B_y$ along the satellite track represent downward currents in this case, and negative gradients represent upward currents. In between the apparent large-scale upward and downward current sheets, highly fluctuating magnetic perturbations up to 800 nT, which challenge the maximum slew rate of MGF, are observed within the ion heating region. Fourier analysis of this time series indicates a broadband-emission feature in the frequency range from 0 Hz to 80 Hz (not shown here), although this may be contaminated by poor instrumental reconstruction of the large, rapid perturbations observed. These turbulent magnetic fields suggest plasma wave activity in the ion heating region. The overall apparent upward currents inside the fluctuating magnetic perturbation region are correlated with active and small-scale auroral ray structures. Whether it is upward or downward current that is associated with the ion heating we are observing is not clear in this region, since the assumption of large-scale infinite current sheets breaks down. 
 
\begin{figure}[t!]
\includegraphics[scale=0.5,width=8cm]{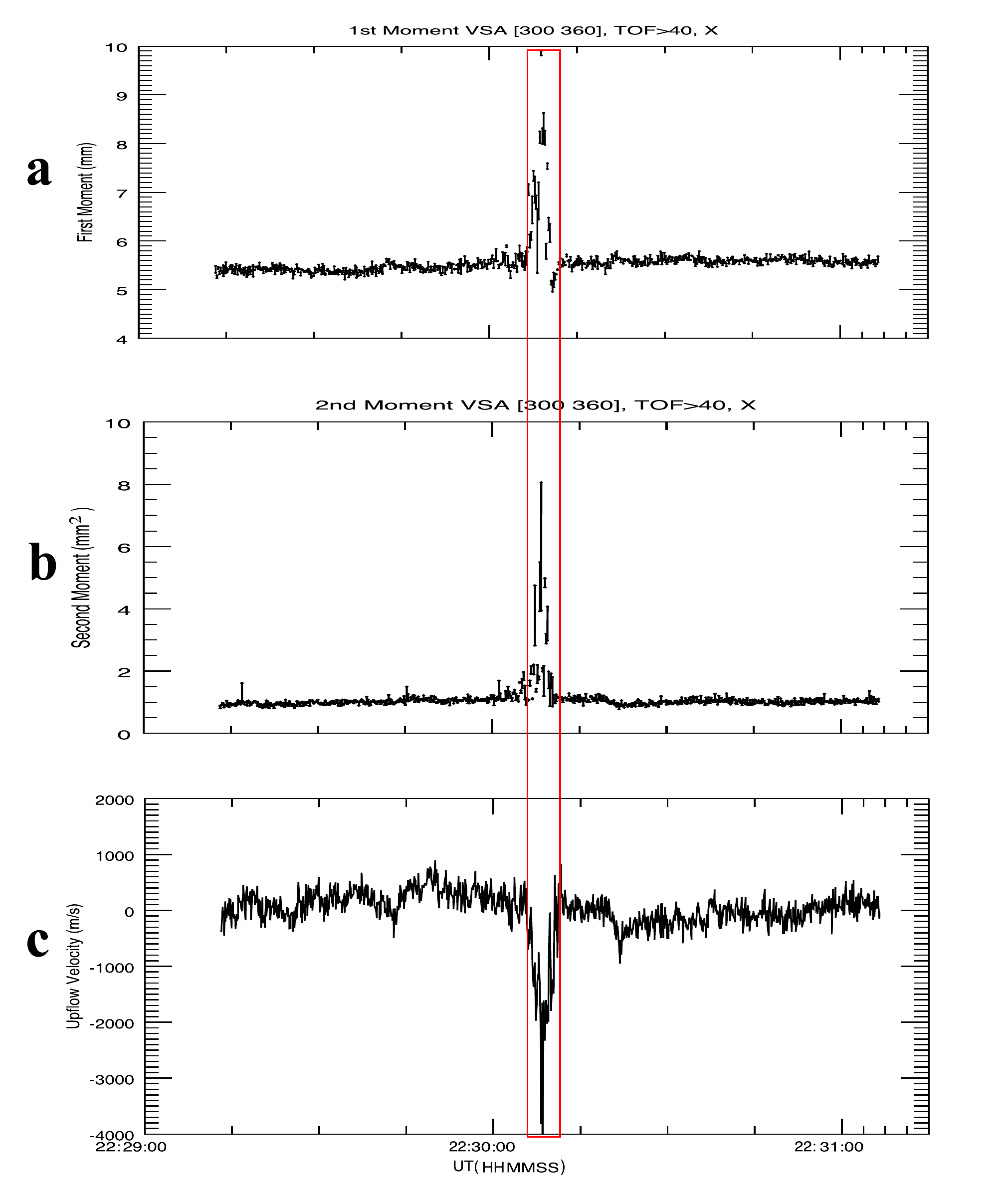}
\caption{Summary plot from the IRM observations of the ion heating and downflow event shown in Figure~\ref{firstsummary}.  a) First moment of the core ion distribution in the direction perpendicular to $\vec{B}$ (in units of mm in the detector plane), representing ion flow enhancement in the ram direction. b) Second perpendicular moment (in units of mm$^{2}$) of the ion distributions, indicating perpendicular ion temperature increases. The observed significant enhancements in average ion bulk flow velocities and perpendicular ion temperatures within the ion heating region (red box region) near 22:30:09 UT are consistent with the SEI measurements as shown in Figure~\ref{firstsummary}. c) Field-aligned ion bulk flow velocities (55 samples per second) from IRM. Ion downflows of over 3 km/s are correlated with ion temperature increases.}
\label{irm}
\end{figure}

 Figure~\ref{irm} presents the measurements from the IRM instrument during the same period. Figure~\ref{irm}a shows the first perpendicular (to $\vec{B}$) moments of ion distributions, representing average perpendicular (to $\vec{B}$) ion bulk flow velocities. Figure~\ref{irm}b displays the second perpendicular (to $\vec{B}$) moments of the ion distributions, indicating perpendicular ion temperature increases. The first moment is shown in units of mm (in detector units), and the second moment is in mm$^{2}$. The observed significant enhancements in average bulk flow ion velocities and perpendicular ion temperatures within the ion heating region near 22:30:09 UT (red box region) are consistent with the SEI measurements as shown in Figure~\ref{firstsummary}a. Figure~\ref{irm}c shows the field-aligned ion bulk flow velocities from IRM, which are calculated every 0.018 second. Again, ion downflows of over 3 km/s (potentially overestimated in a manner similar to SEI estimates) are correlated with ion temperature increases. IRM TOF measurements show that O$^{+}$ ions are the major (approximately 99\%) ion species in this event.

\begin{figure*}
\includegraphics[scale=.4, width=16cm]{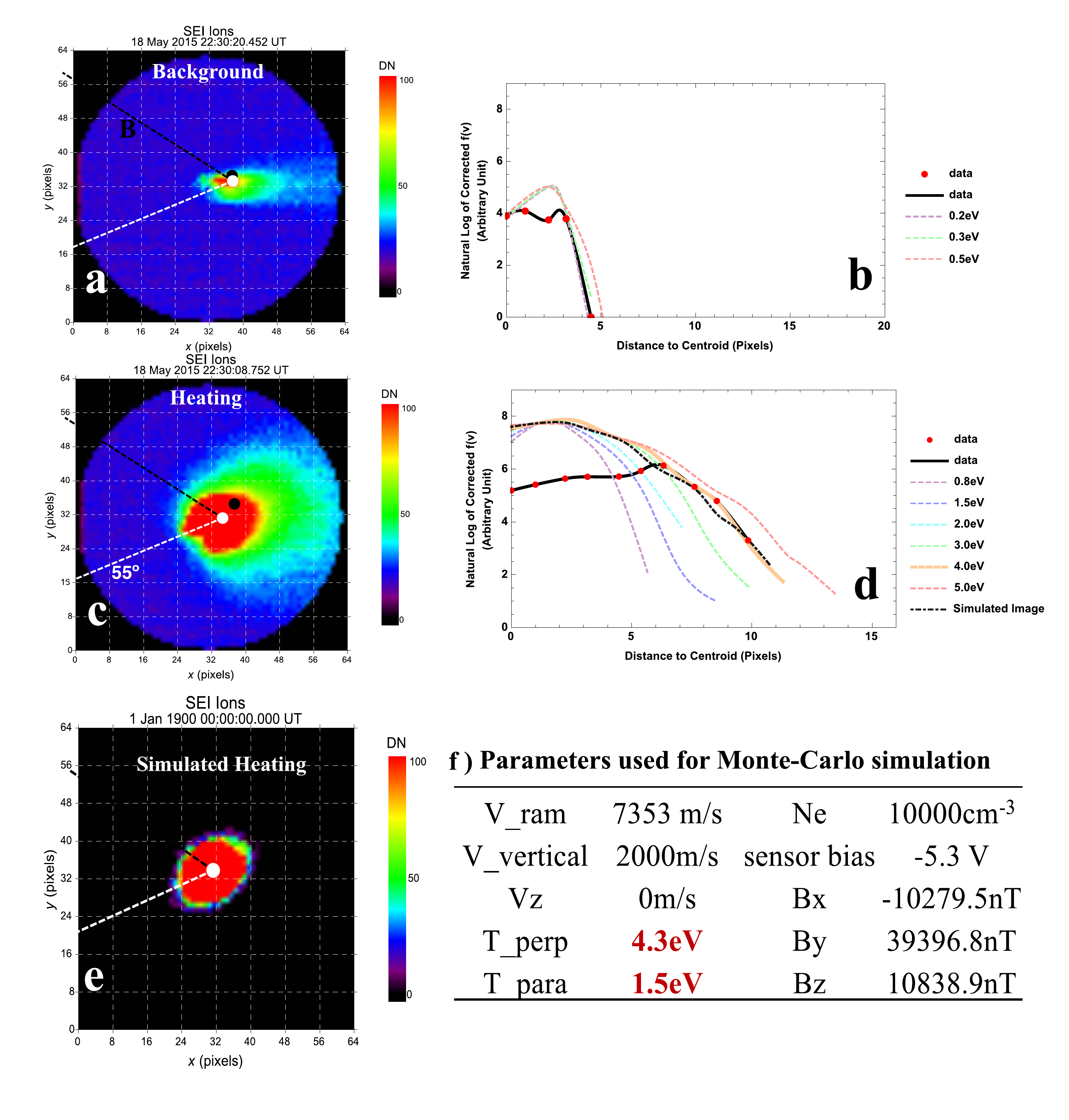}
\caption{Calibration of ion temperature measurements through forward modeling of the flight images and matching the simulated ion distribution function slopes with observations. a) A typical background flight image measured outside the heating region. b) Comparisons between simulated distribution functions with different Maxwellian temperatures, and the measured one, which is best fitted with a Maxwellian ion temperature of 0.2 eV. c) Raw SEI image during the heating period. Ion energies increase with radius, reaching 230 eV at the edge of the detector. d) Comparisons between simulated distribution functions with different Maxwellian temperatures, and the measured one. The heating-time ion distribution function is best fitted with a Maxwellian temperature of 4 eV. The discrepancy between the simulated distribution function curve and the measured one is mainly due to gain variations and stray fields affecting low-energy ions ending up at the image center. e) Forward-modeled image that approximately matches the flight image. f) Parameters used for this Monte-Carlo simulation. The perpendicular ion temperature is 4.3 eV. The parallel ion temperature used is 1.5 eV.}
\label{firstvalid}
\end{figure*}

 Figure~\ref{firstvalid} represents a calibration of ion temperature measurements through forward modeling of the flight images and matching the simulated ion distribution function slopes with observations. Figure~\ref{firstvalid}a shows a typical background flight image measured outside the heating region. The white dot represents the image centroid position, while the black dot represents the image center position. The black line indicates the direction of the magnetic field projected onto the detector plane. The ion distribution function profile, measured in terms of data number (arbitrary intensity units at the CCD output) versus distance to the centroid position, is obtained for a particular pitch angle direction (55\deg~, indicated by the white line) that has the most evident width increase. Figure~\ref{firstvalid}b shows comparisons between simulated distribution functions and the measured one. The red dots and the black line are observed data. The lines with color represent simulated data. The background distribution function slope is best fitted with a Maxwellian ion temperature of 0.2 eV. 
 
 Figure~\ref{firstvalid}c shows the raw SEI image during the heating period. The ion signals in red color, which are in general less than 100 eV, are used for the distribution function fitting. The heating-time ion distribution is clearly larger in width compared to the background flight image. Figure~\ref{firstvalid}d shows the comparisons between simulated distribution functions with different Maxwellian temperatures, and the measured one during the heating period. The ion distribution function is best fitted with a Maxwellian temperature of 4 eV. This result is consistent with the perpendicular ion temperature derived from the FWHM measurement as shown in Figure~\ref{firstsummary}a. The discrepancy between the simulated distribution function curve and the measured curve mainly lies within $\sim$5-pixel range from the center position. This is due to non-ideal effects including gain variations and stray fields affecting ions having low energies. The ion signal is not saturated near the image center. Figure~\ref{firstvalid}e displays the forward-modeled image that approximately matches the flight image. The parameters used for this Monte-Carlo simulation are given in Figure~\ref{firstvalid}f. The perpendicular ion temperature is 4.3 eV, the same as that measured from the FWHM variations. However, the parallel ion temperature is 1.5 eV instead of 3 eV as indicated by the parallel FWHM variations (Figure~\ref{firstsummary}b). 
 
\begin{figure*}
\includegraphics[height=9cm,width=15cm]{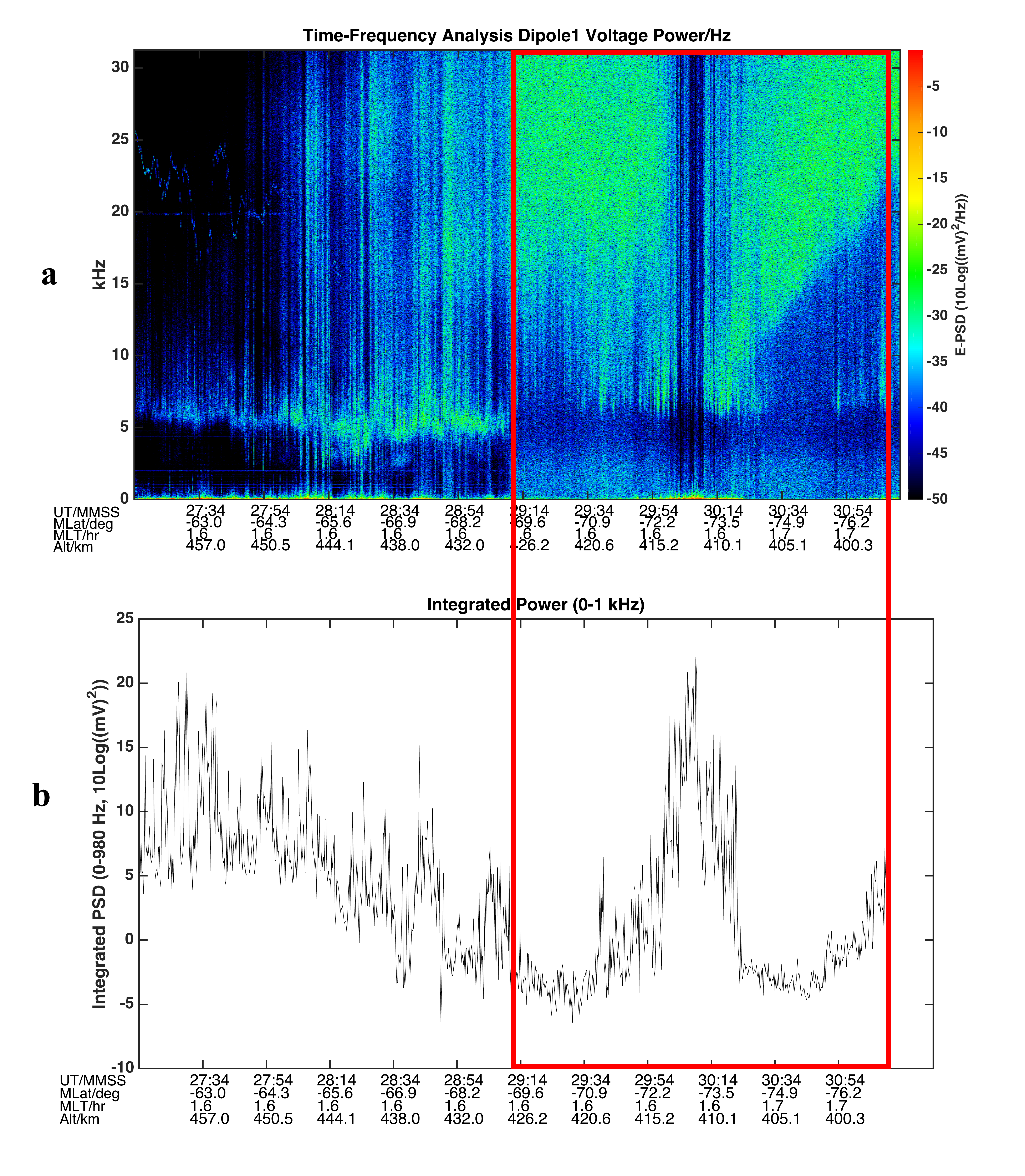}
\caption{Observations of wave electric fields from the RRI instrument. a) Time-frequency spectrogram of the wave electric fields measured from one dipole antenna. The red shaded region indicates the period when the SEI instrument is operating simultaneously with RRI. A so-called ``saucer" emission signature is identifiable between 22:30:14 and 22:30:54 UT. BBELF wave emissions near the vertex of the VLF saucer, extending from a few Hz up to approximately 1 kHz, are observed near 22:30:09 UT when the SEI instrument observes strong ion heating signatures. b) PSD values integrated over frequency from DC up to 1 kHz throughout the orbit. It is clear that the ion heating event observed by the SEI instrument is correlated with strong BBELF wave emissions.}
\label{rri}
\end{figure*}

 Figure~\ref{rri} presents the observations from the RRI instrument, which characterize the wave electric fields in the frequency range from a few Hz up to approximately 30 kHz. Figure~\ref{rri}a shows the time-frequency spectrogram result of the wave electric fields measured from one dipole antenna. Measurements from the other dipole demonstrate very similar spectral features and are therefore not repeated here. The red shaded region indicates the period when the SEI instrument is operating simultaneously with the RRI. From 22:29:13 to 22:31:11 UT, RRI shows auroral hiss emissions which have a clear cutoff frequency near 6 kHz. A so-called VLF ``saucer" emission signature \citep{smith1969,james1976} is identifiable between 22:30:14 and 22:30:54 UT. Note that the VLF wave power is depleted in some regions near the vertex; this is due to saturation of the dipole electric field measurement (up to 8 mV). It is reasonable to believe that there is even larger hiss wave power inside the depletion region near the vertex \citep{lonnqvist1993,ergun2001}. The enhancement in the low-frequency wave power near the depletion region is consistent with the fact that coherent low-frequency wave structures are observed in the electric field time series without saturation. 

 Near 22:30:09 UT when the SEI instrument observes strong ion heating signatures, there are intense BBELF wave emissions extending from a few Hz up to approximately 1 kHz. PSD intensities reach over 1 mV$^{2}$/Hz near DC frequency and gradually decrease to 10$^{-2}$ mV$^{2}$/Hz near 500 Hz. These BBELF waves occur near the vertex of the VLF saucer, which suggests the process is related to the return current (downward current) region \citep{lonnqvist1993,ergun2003}. Figure~\ref{rri}b presents the PSD values integrated over the frequency bins from DC up to 1 kHz throughout the orbit. It is clear that the ion heating event observed by the SEI instrument is correlated with strong BBELF wave emissions. 

\begin{figure*}
\includegraphics[height=9cm,width=15cm]{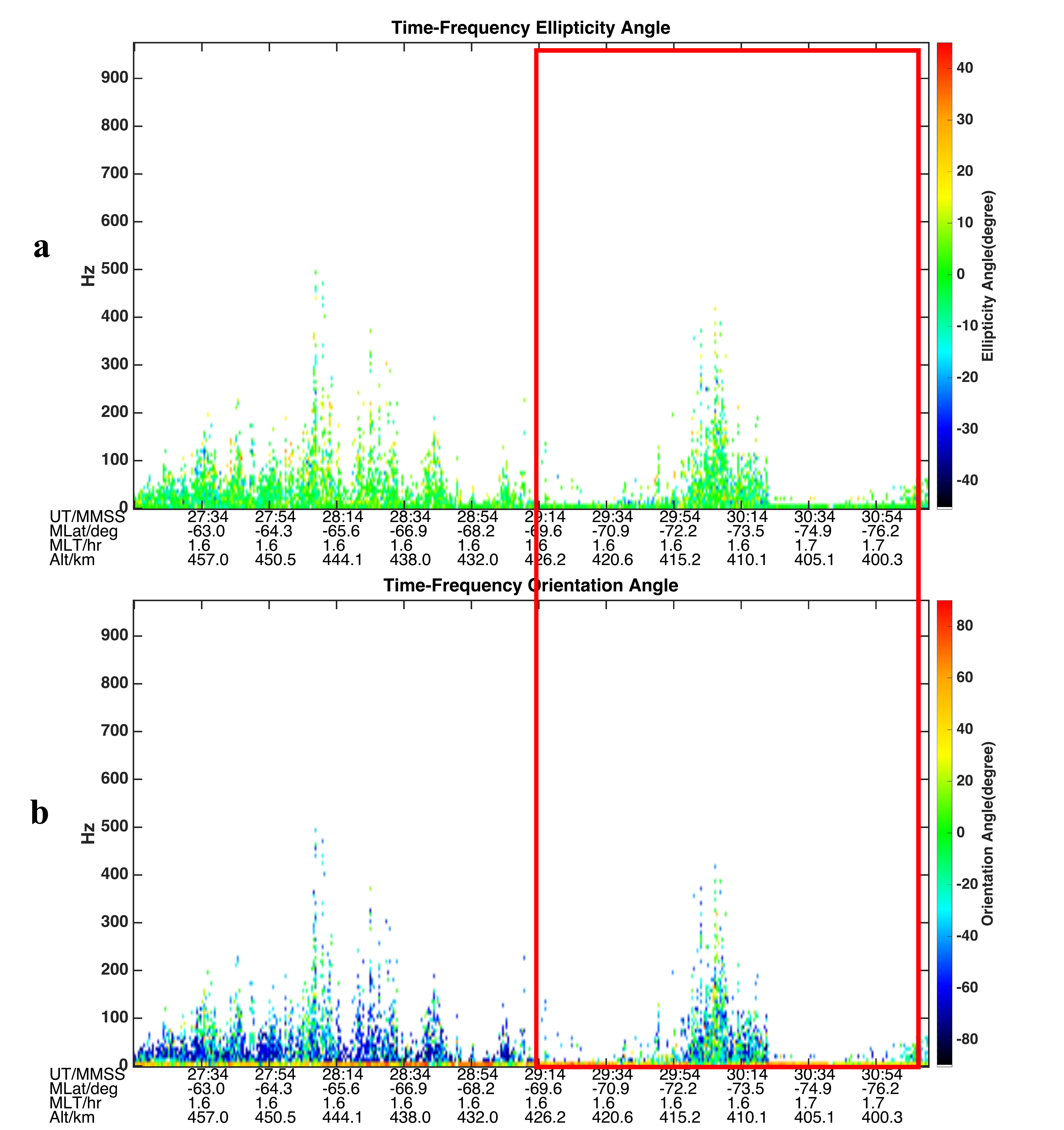}
\caption{Ellipticity angle (a) and orientation angle (b) spectrograms with respect to frequency and time for the nightside auroral event. The frequency range is from a few Hz up to approximately 1 kHz. We apply a threshold value in PSD when calculating the ellipticity and orientation angle so as to exclude background noisy angles, which are unreliable due to the relatively small values (1/150 times the maximum PSD value) used in the tangent function. Therefore PSD in the blank white region is relatively small and negligible. Inside the ion heating region (red box region), the ellipticity angles of the measured BBELF waves approach 0\deg. This indicates that the BBELF waves are in general linearly polarized. The orientation angles are roughly -45\deg~throughout the BBELF emission region, meaning that the linearly polarized electric fields are approximately perpendicular to the background magnetic field. }
\label{polarization}
\end{figure*}

 Figure~\ref{polarization} shows the ellipticity angle and orientation angle spectrograms from a few Hz up to approximately 1 kHz. Inside the ion heating region, the ellipticity angles of the measured BBELF waves approach 0\deg. This indicates that the BBELF waves are in general linearly polarized. The orientation angles are roughly -45\deg~throughout the BBELF emission region, meaning that the linearly polarized electric fields are nearly perpendicular to the background magnetic field. However, considering that the magnetic field direction is approximately 105\deg~with respect to the dipole plane normal direction in this case, one could argue that a circularly polarized electromagnetic wave that propagates roughly parallel to the dipole plane may also manifest itself as linearly polarized in our cross-dipole plane. This turns out to be less likely due to the fact that the RRI does observe right-hand circularly polarized chorus waves in some cases where the magnetic field direction is 99\deg~away from the dipole plane normal. The ellipticity angles ($\chi$) are close to 20\deg~and are in sharp contrast to the linearly polarized waves in the background (not shown here). This suggests that the RRI instrument can measure circularly polarized whistler waves if the angle between the magnetic field direction and the dipole plane is large enough, in which case the RRI instrument is measuring the projection of the polarization ellipse onto the dipole plane.
 
 
Near-infrared auroral images measured from the FAI instrument onboard are included in the Supplementary Information as a movie. The magnetic footprint of the satellite is located outside but near the field view of the camera. The coordinate grids shown along with the movie frame are in geographic coordinates. The magnetic footprint traverses geomagnetic latitudes from -70\deg~to -75\deg~during the movie. From 22:29:45 to 22:29:52 UT, the spacecraft encountered relatively stable multiple auroral arcs, during which time the MGF observed a relatively uniform upward current sheet as shown in Figure~\ref{firstsummary}d. From 22:30:05 to 22:30:10 UT, the magnetic footprint came across a region of very active auroral rays, which overlap with the ion heating region as identified previously in Figure~\ref{firstsummary}a. The strongest perpendicular ion temperature increase of up to 4.3 eV is associated with these highly active auroral structures.

\subsection{Example in the Dayside Cleft }

Another example of ion heating took place in the dayside cleft region on May 16, 2015 at an altitude of 357 km in the southern hemisphere. From 21:19:30 to 21:21:40 UT, the e-POP satellite travelled from AACGM latitude -79\deg~to -71\deg~near the magnetic local time (MLT) 15.2 hr. Measurements from the SEI and MGF instruments are available during this time. SEI was operating on Normal mode with $V_{inner}$ = -502 V, indicating the same ion energy range in the SEI image as in the previous example. From 21:20:07 to 21:20:23 UT, the SEI instrument observed a large region of ion heating that persists nearly 15 s along the satellite track, corresponding to a spatial region nearly 120 km across. The most intense perpendicular ion temperature increase goes up to 4.5 eV based on the distribution function analysis applied to the SEI ion images. The special aspect of this event is that there are two ion populations with different energies and counterstreaming field-aligned ion bulk flow velocities resolved from the SEI measurements inside the ion heating region. The core ions, with energies up to 20 eV, are flowing downward while the tail ions with higher energies are flowing upward along the magnetic field line. 

\begin{figure}[h]
\includegraphics[height=9cm,width=8cm]{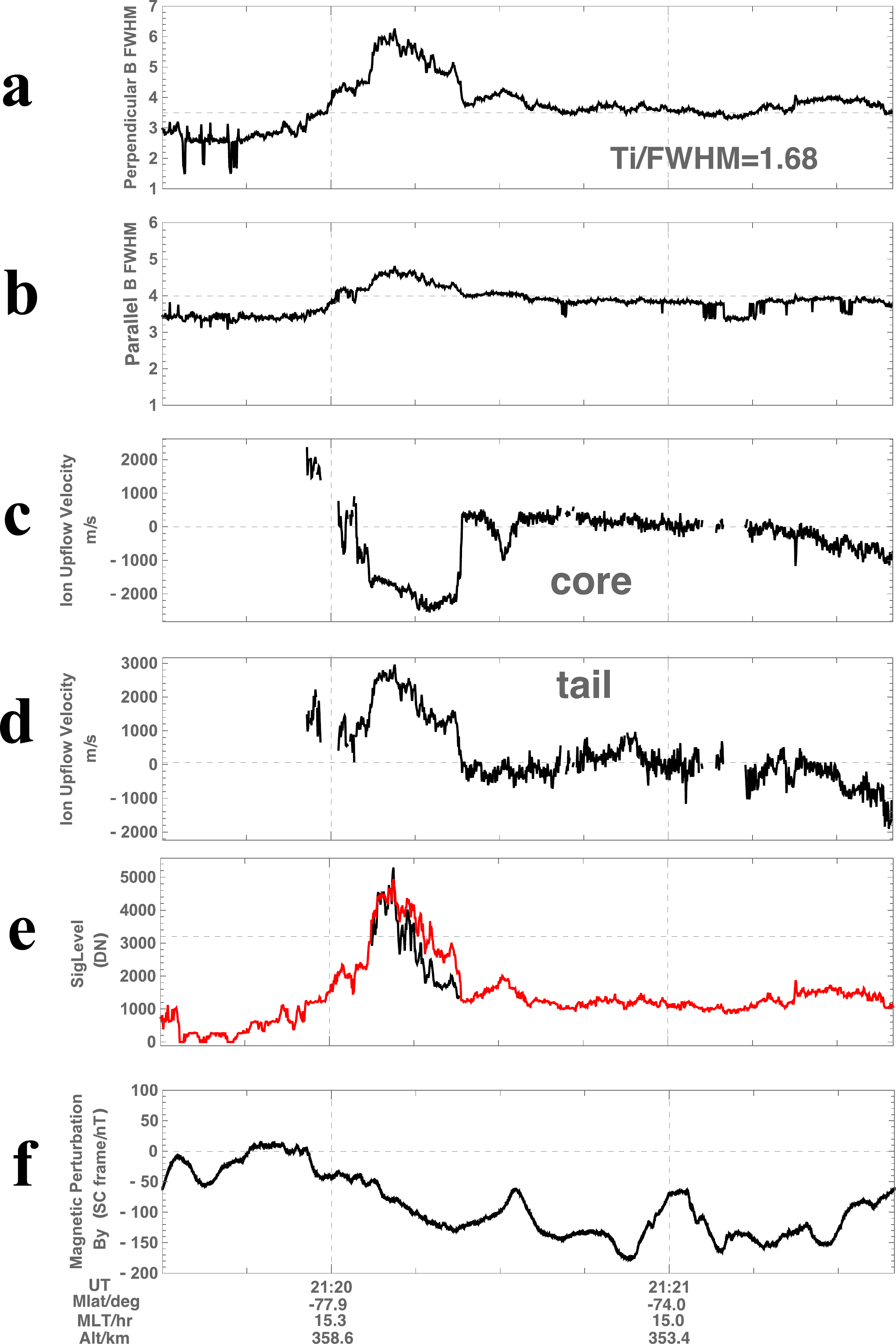}
\caption{e-POP SEI (ion mode) and MGF observations of the ion heating event in the dayside cleft on 16 May 2015 at an altitude of 357 km. a) FWHM variations perpendicular to $\vec{B}$. The enhancement near 21:20:14 UT represents a perpendicular ion temperature increase of up to 4.2 eV. b) FWHM profile parallel to $\vec{B}$. The relatively small increase in the FWHM indicates parallel ion temperature increase of roughly 1 eV. c) Field-aligned flow velocities of the core (\textless 20 eV) ions. In this event core ions are flowing down the field line with velocities of over 2 km/s inside the heating region. d) Field-aligned ion flow velocities of the tail ion population. The high-energy ions are flowing upward with velocities of approximately 3 km/s inside the heating region. e) Ion signal level, approximately proportional to ion flux, for the core (red) and tail (black) ion populations. f) Magnetic field perturbations of the $\Delta B_y$ component in the spacecraft frame. This ion heating event occurs in the return current region.}
\label{secondsummary}
\end{figure}

Figure~\ref{secondsummary} summarizes the observations from the SEI and MGF instruments, outlining the ion temperatures, field-aligned ion bulk flow velocities and magnetic field measurements as a function of time. Figure~\ref{secondsummary}a shows the perpendicular (to $\vec{B}$) FWHM variations. Every unit increase in perpendicular FWHM under Normal mode corresponds approximately to a 1.6 eV increase in the perpendicular ion temperature. The enhancement near 21:20:14 UT represents a perpendicular ion temperature of up to 4.2 eV. Figure~\ref{secondsummary}b displays the parallel FWHM profile, which has relatively small enhancements during the ion heating period. Such an increase in the FWHM indicates a roughly 1 eV parallel ion temperature increase. 

Figure~\ref{secondsummary}c shows the field-aligned bulk flow velocities of the core ions. The contribution of tail ions, with energies greater than near 20 eV, is excluded during the analysis. The bulk flow velocities are calculated every 0.1 s. Ion images with insufficient ion signal, or the centroid positions which are very close to the imager center position (less than 0.4 pixel) are deemed bad data and excluded from the calculation. In this event core ions are flowing downward along the field line with velocities of over 2 km/s inside the ion heating region. Figure~\ref{secondsummary}d shows the field-aligned ion bulk flow velocities when we include the tail ion population in the ion image moment calculations. The high-energy ions are flowing upward with bulk velocities of near 3 km/s inside the ion heating region. 

Figure~\ref{secondsummary}e presents the ion signal level, representing approximate ion fluxes, for the core and tail ion population. During the ion heating period, the tail ion flux increases to a level comparable to the core ion flux. This suggests that the tail ion population might originate from ionospheric regions rather than coming from much higher altitudes in the magnetosphere, where plasma densities are much smaller than in the ionosphere. The last panel in Figure~\ref{secondsummary} shows the magnetic field perturbations in the $B_y$ component in spacecraft frame. When the satellite is travelling from high latitudes to lower latitudes, the negative gradients of the $B_y$ perturbations represent downward currents. Therefore this ion heating event occurs in the return current region. 

\begin{figure*}
\includegraphics[height=10cm,width=16cm]{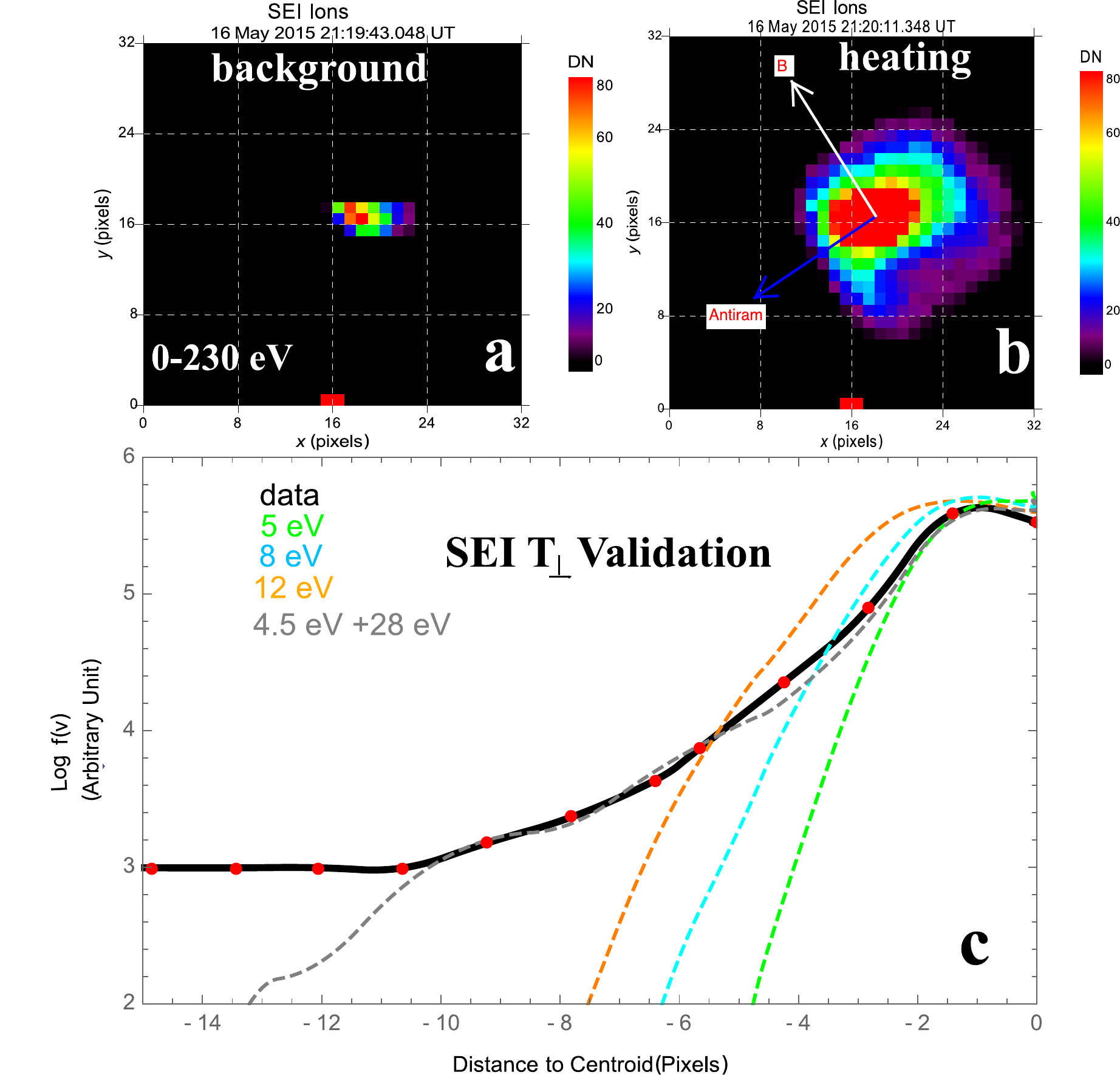}
\caption{Forward modeling the ion distribution function slopes for the cleft example with different Maxwellian ion temperatures based on simulation. a) Background flight image. b) Heating-time flight image. Ions in the red-color region represent the core ion population, while the ions in green and blue are defined as high-energy ions (tail ions). The magnetic field direction projected onto the SEI image along with the satellite ram direction is indicated in the image. c) Perpendicular ion distribution function fitting for heated image. The red dots and black line are the measured perpendicular ion distribution function data. The green, cyan, and yellow dashed lines represent ion distribution functions with 5 eV, 8 eV, and 12 eV Maxwellian temperatures from the simulation. This distribution function slope can only be matched by two ion populations with different Maxwellian temperatures, as indicated by the gray dashed line. The core ion temperature is 4.5 eV and the tail ion temperature is near 28 eV. }
\label{secondvalid}
\end{figure*}

The heating-time ion temperature is also checked through forward modeling the ion distribution function slopes with different Maxwellian ion temperatures based on the Monte-Carlo instrument simulation. Figure~\ref{secondvalid} shows a background flight image (Figure~\ref{secondvalid}a) and the heating-time flight image (Figure~\ref{secondvalid}b) along with the perpendicular ion distribution function fitting result (Figure~\ref{secondvalid}c). As before, with the deflection voltage $V_{inner}$ = -502 V, the sensor measures ions with energies up to $\sim$230 eV (at the edge of the image). However, most of the ions detected in this event have energies less than 100 eV. Ions in the red-color region represent the core ion population, while ions in green and blue are defined as high-energy ions (tail ions) as mentioned previously. The magnetic field direction projected onto the SEI image along with the satellite ram direction is indicated in the image. 

In Figure~\ref{secondvalid}c, the red dots and black line are the perpendicular ion distribution function data measured in Figure~\ref{secondvalid}b. This distribution function slope can only be matched by two ion populations with different Maxwellian temperatures, as indicated by the gray dashed line. The core ion temperature is 4.5 eV and the tail ion temperature is near 28 eV. Such a core ion temperature of 4.5 eV is consistent with the result from the FWHM measurement as presented in Figure~\ref{secondsummary}a. We should mention here that \citet{garbe1992} made a similar observation from TOPAZ 2 sounding rocket. They found the core ions with 3 eV ion temperature are flowing downward with velocities between 3-5 km/s and the tail ions are going upward with a temperature of tens of eV. The counterstreaming ions with different temperatures have also been observed from the SIERRA sounding rocket \citep{lynch2007}, except that the core ions are flowing upward while the high-energy ions are flowing downward in their study.

\subsection{Statistics: 24 Events }

\begin{figure*}
\includegraphics[height=8cm, width=15cm]{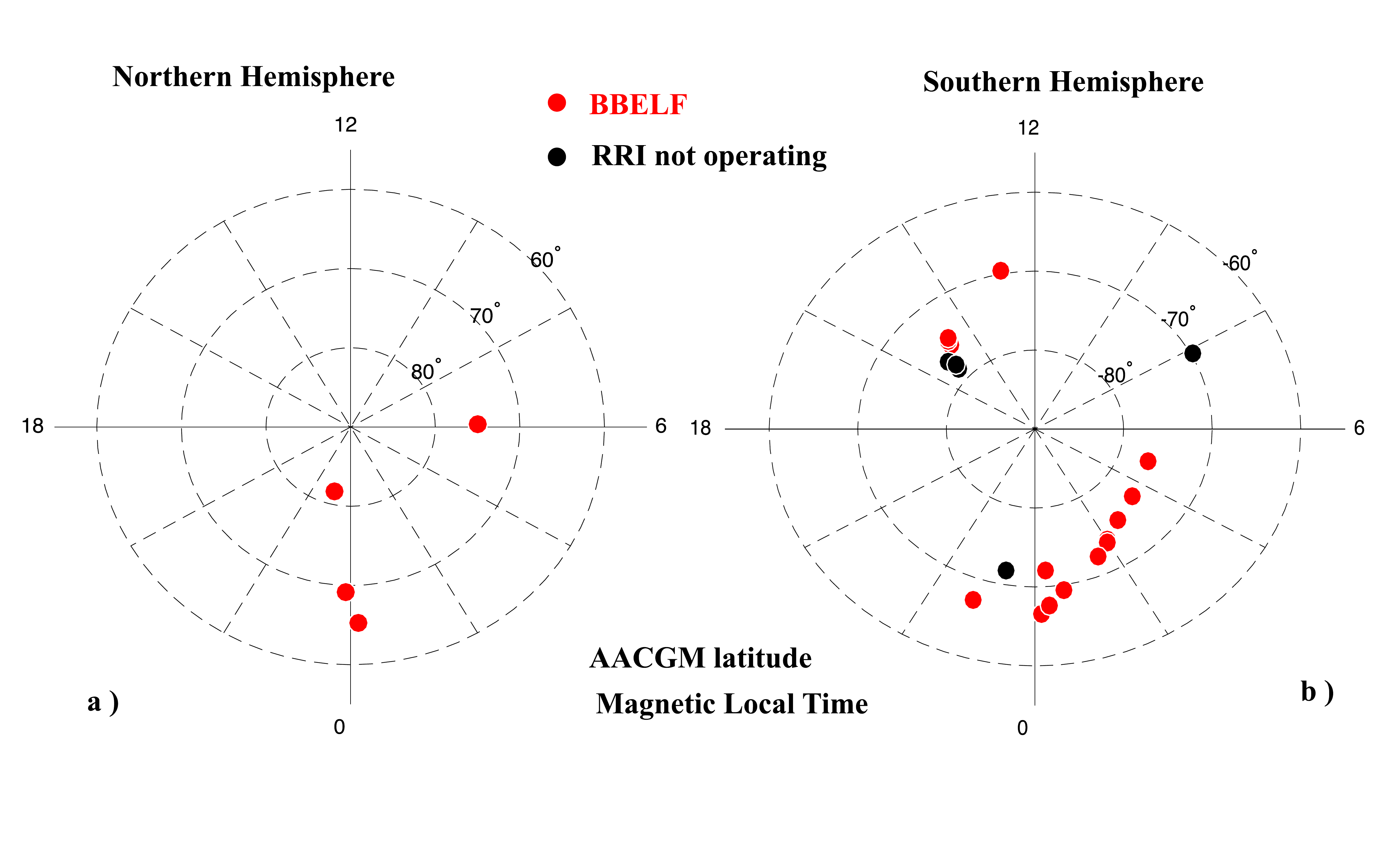}
\caption{Statistical locations of 24 ion heating events on the AACGM latitude and MLT polar plot in the southern (a) and northern (b) hemisphere respectively. The majority of the events (19 out of 24) occurred in the nightside auroral region in the southern hemisphere. Red circles indicate that the RRI is observing BBELF wave emissions simultaneously with ion temperature enhancements. There are 18 ion heating events that are accompanied by the BBELF waves. Black circles indicate that the RRI is not operating at the same time. This map indicates that the ion heating events are well correlated with the BBELF wave emissions.}
\label{map}
\end{figure*}

From the period March 2015 to March 2016, we found in total 17 orbits demonstrating in total 24 ion heating events observed from the SEI instrument. The majority of them have simultaneous observations from the IRM, the MGF and the FAI to provide complementary perspectives on the ion heating events. Figure 9 presents the locations of these ion heating events on the AACGM latitude and MLT polar plot in the southern (Figure~\ref{map}a) and northern (Figure~\ref{map}b) hemisphere respectively. The majority of the events (19 out of 24) occurred in the nightside auroral region in the southern hemisphere. Note that this statistical map cannot fully represent the occurrence rate of ion heating. The smaller number of events in the northern hemisphere and in the dayside cleft region, for example, is due to poor coverage of the operation of the instruments as well as bad data during some operations. 

The key point of this map is to show correlations of ion heating events with observations of BBELF waves from the RRI instrument. There are 18 ion heating events that are accompanied by the BBELF waves. Such a high correlation of ion heating events with BBELF waves has been reported previously near 1700 km altitude from Freja satellite observations \citep{knudsen1998}, but not in the altitude range that our study is focusing on. The BBELF waves measured in our cases are consistently linearly polarized and have orientation angles roughly perpendicular to the local magnetic field. Many events have concurrent magnetic field fluctuations measured from MGF, showing wave structures in the frequency range from near DC up to above local $O^{+}$ cyclotron frequency, such as the one demonstrated in Figure~\ref{firstsummary}d. Therefore there are potentially both electromagnetic and electrostatic wave modes in the BBELF waves.

Table~\ref{table} summarizes the characteristics of the observed 24 ion heating events in terms of the orbit time, field-aligned flow velocity pattern, FAC pattern, whether the RRI is observing BBELF waves, whether the FAI is observing aurora, altitude, and geomagnetic activity. With respect to the FAC pattern, only 2 out of 24 events show upward currents associated with the ion heating. The ion heating events are more likely to be correlated with downward currents, although several events are within the transition region between the upward and downward current region. Most of the events occur in the altitude range from 325 km to 730 km, and are associated with auroral activity. 

Of particular interest in the table is that 17 out of 24 ion heating events have core ion bulk downflow rather than upflow. This is contrary to what would be expected from the upward mirror force acceleration of heated ions. It is noted that we can only validate ion temperatures from orbits that are shown in green color. The SEI instrument was running on a very low deflection voltage (-26 V) for many events and the results from the simulation failed to reproduce the flight data. 

\begin{figure*}
\includegraphics[height=10cm,width=\textwidth]{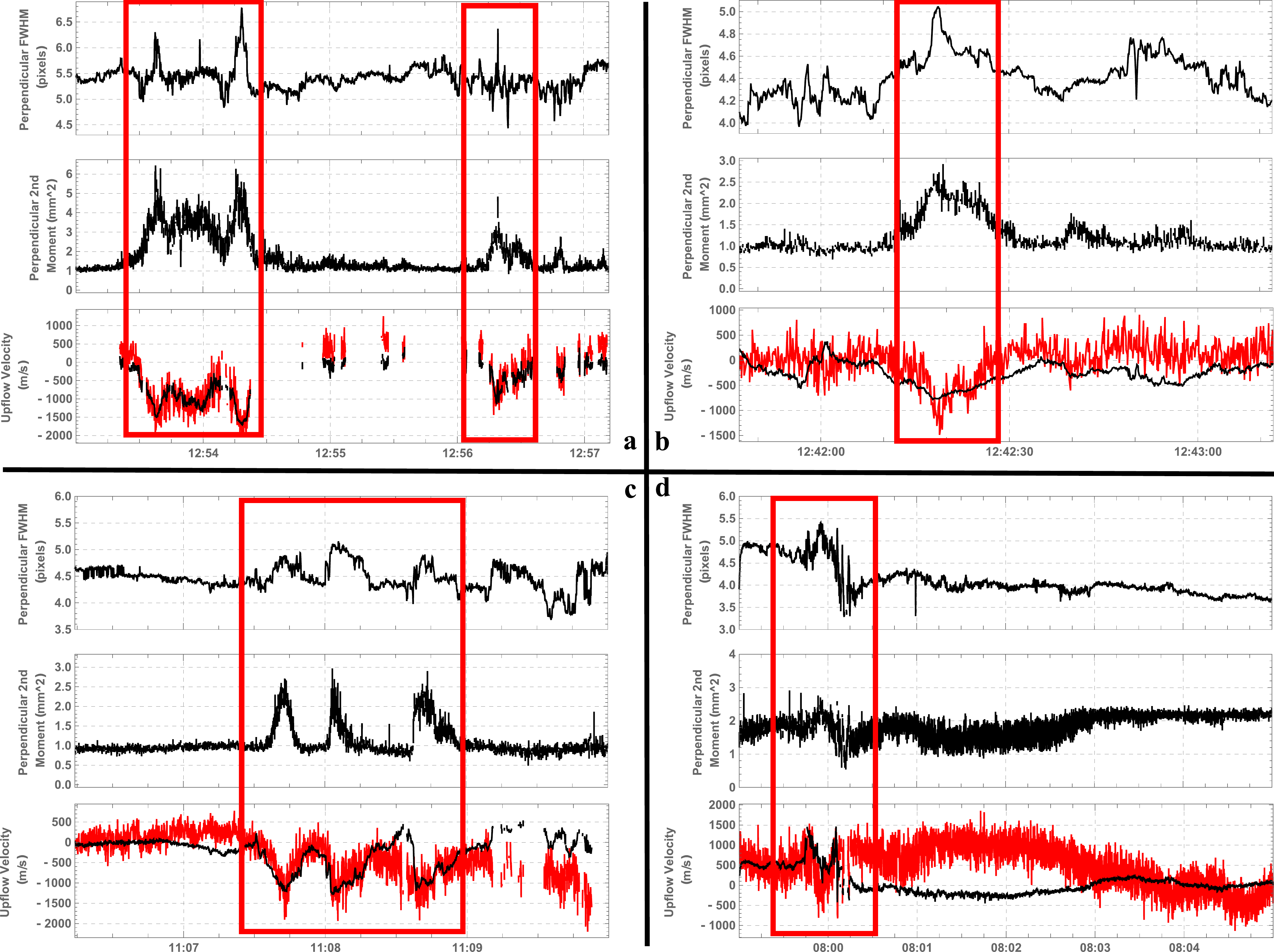}
\caption{Examples of correlation between ion heating and ion downflow. The four subfigures demonstrate observations from four events, on June 8th, 9th, 16th, and September 19th, 2015, respectively. The top panel of each subfigure displays the perpendicular FWHM variations from the SEI observations, indicating perpendicular ion temperature increases. The middle panels show the perpendicular second moment profile from the IRM measurements, representing perpendicular ion temperature variations as well. The bottom panels of these subfigures display field-aligned ion bulk flow velocities calculated approximately every 0.01 second from both the SEI (black) and IRM (red) instruments. Consistently throughout all these events and from both instruments, the observed perpendicular ion temperature increases are well correlated with the ion downflow velocities.}
\label{excerpt}
\end{figure*}

Figure~\ref{excerpt} presents several ion heating events exemplifying the correlation between the ion heating and ion downflow. The four subfigures demonstrate the perpendicular ion temperatures and field-aligned flow velocities measured from both the SEI and IRM instruments from four events, on June 8th, 9th, 16th, and September 19th, 2015, respectively. There are 7 ion heating events in total in this figure. The top panels of these subfigures display the perpendicular FWHM variations from the SEI observation, indicating perpendicular ion temperature increases. The middle panels show the perpendicular second moment profile from the IRM measurements, representing perpendicular ion temperature variations as well. The bottom panels of these subfigures display the field-aligned ion flow velocities calculated every 0.01 second from both the SEI (black) and IRM (red) instruments. Consistently throughout all these ion heating events and from both instruments, the observed perpendicular ion temperature enhancements are well correlated with the ion downflow velocities. One particularly intriguing feature in Figure 10d is that core ion upflows are observed adjacent to the ion heating region. Once the satellite enters into the ion heating region, the core ions are observed to flow downward. 

\begin{table}
\caption{Characteristics of the observed 24 ion heating events in terms of the orbit time, field-aligned flow velocity pattern, FAC pattern, wave activity, auroral activity, altitude, and geomagnetic activity. Only 2 out 24 events show upward currents associated with the ion heating. The ion heating events are more likely to occur in downward currents, albeit several events are within the transition region between the upward and downward current region. ``N/A", i.e., not applicable is used for one event that has no MGF data and another that shows no significant magnetic perturbations since the satellite is travelling nearly parallel to an auroral arc. All events occur in the altitude from 325 km to 730 km, and most are associated with auroral activity. Of the 24 ion heating events, 17 are correlated with core ion bulk downflow rather than upflow.}
\includegraphics[scale=.25]{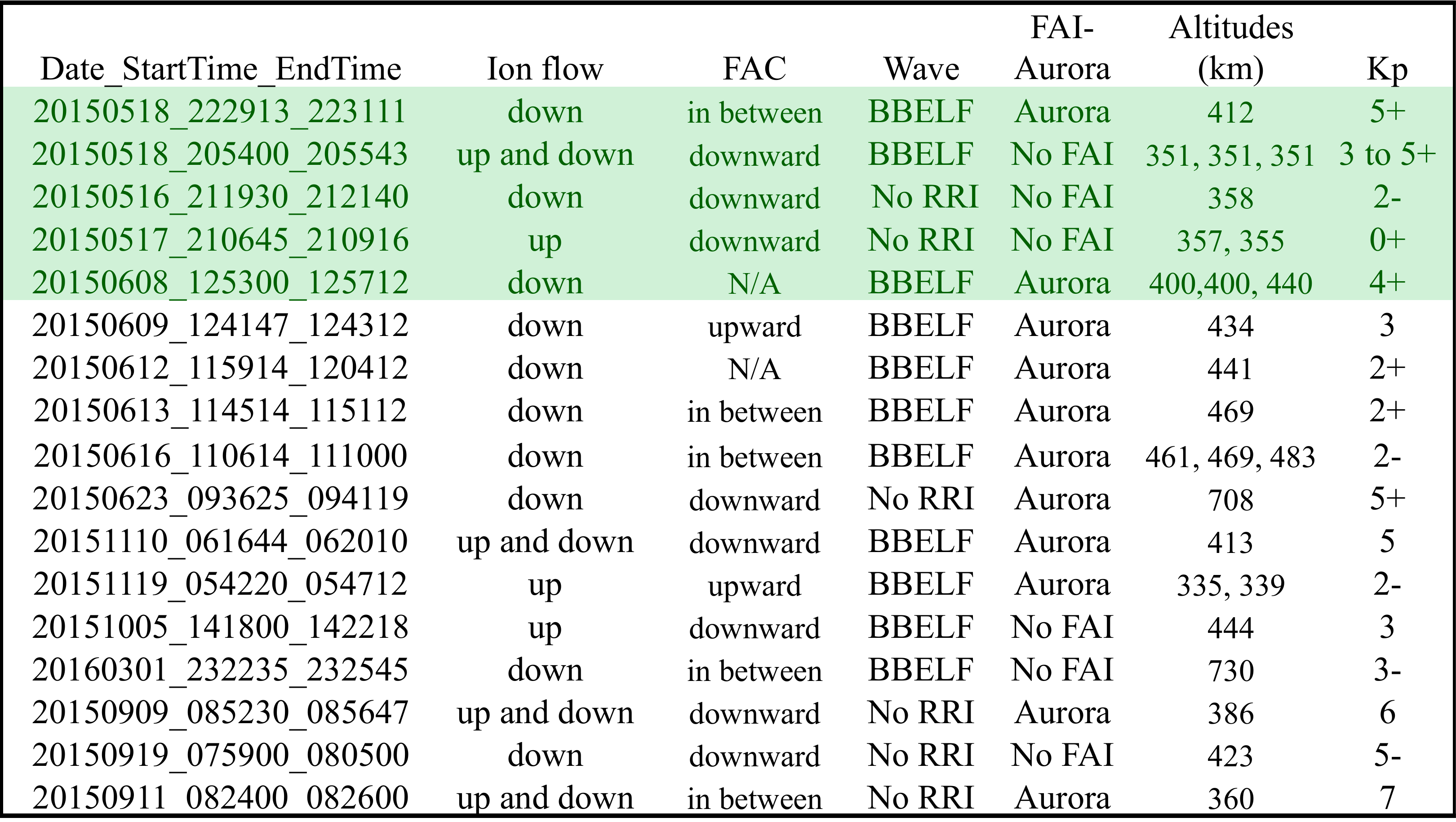}
\label{table}
\end{table}

\section{Discussion}

As discussed in the Introduction section, two mechanisms, namely frictional heating and wave-ion heating, are commonly invoked to explain perpendicular ion temperature increases near the topside ionosphere. Despite the fact that frictional heating occurs quite frequently at auroral latitudes near the topside ionosphere due to large convection electric fields, the amount of thermal energy that can be drawn from that mechanism is limited. Based on a time-dependent gyro-kinetic model of the high-latitude F-region response to frictional heating, \citet{loranc1994} found that transient values of the perpendicular ion temperature due to frictional heating can only reach 2 eV in extreme cases. Some other frictional heating models near the topside ionosphere \citep{heelis1993, wilson1994, liu1995} rarely showed ion temperature increases greater than 1 eV. If one completely converts the kinetic energy of O$^{+}$ ions flowing with a 4 km/s horizontal velocity, a value which is from previous studies rarely observed near the topside ionosphere, into ion thermal energy, the resulting ion temperature is only 1.3 eV. 

However, as shown in the two ion heating examples discussed in the Data section, the perpendicular ion temperatures reach as high as 4.5 eV (Figure~\ref{firstsummary}a, Figure~\ref{firstvalid}, Figure~\ref{secondsummary}a, Figure~\ref{secondvalid}). Therefore the ion heating events that we observed cannot be explained by frictional heating. Rather we argue that wave-ion heating plays a dominant role in these ion heating events. The correlation between ion temperature enhancements and BBELF waves simultaneously observed in most of our events, as shown in Figure~\ref{rri} and Figure~\ref{map}, lends solid support to this wave-ion heating interpretation. It is also noted that the parallel ion temperatures are less than half of the observed perpendicular ion temperatures, which is consistent with both the frictional heating and wave-ion heating mechanisms. The increase in parallel ion temperatures may result from isotropizing effects due to collisions with neutrals.

A significant new finding of this study is the observation of wave-particle heating at altitudes well below 450 km, which is the lowest strong heating event reported previously \citep{whalen1978,yau1983}. Our statistical survey with simultaneous low-energy ion and low-frequency wave observations provides empirical evidence of this newly-identified lower boundary (350 km). This is critical to relevant models which study the ion energization processes and the associated field-aligned ion flow near the topside ionosphere \citep{varney2015,burleigh2017}. 

One particular issue that deserves further clarification is whether the observed BBELF waves create the ion heating or whether instead the anisotropic ion distributions resulting from ion heating excite instabilities that generate the waves. It is found that in general the BBELF wave emissions observed along the satellite track have much wider spatial scale than the ion heating regions, such as shown in Figure~\ref{firstsummary}a and Figure~\ref{rri}. These emissions extend over 10 seconds along the orbit, corresponding to 80 km across the magnetic field, while the most intense ion heating occurs within a spatial region only 2 km wide. Also, there are many cases in which the RRI observes BBELF waves but there is no identifiable ion heating signature from either the SEI and IRM instruments (not shown here). This demonstrates that heated ion distributions are not necessary for wave generation, but BBELF waves are present for all the heating events presented here (at least for those for which wave measurements are available). 

In Figure~\ref{rri}, we observe that the BBELF waves are within the vertex region of a VLF saucer. \citet{james1976} suggested that the generation of VLF saucers is associated with an instability from cold ($\textless$ 5 eV) upgoing electron beams interacting with background thermal electrons in the downward current region. This postulate was later confirmed by the Viking and FAST satellites observations at higher altitudes \citep{lonnqvist1993, ergun2001}, who in addition suggested that the upgoing low-energy (near 100 eV) electron beams are accelerated by a downward-pointing electric field in the downward current region. \citet{lonnqvist1993} and \citet{ergun1998} also proposed that these upgoing suprathermal electrons generate so-called ``phase space holes" which in turn are responsible for the excitation of BBELF waves in this region. Based on observations from the GEODESIC sounding rocket above 800 km altitudes coupled with estimations from the WHAMP wave solver, \citet{knudsen2012} proposed that highly localized ELF/VLF wave cavities can be interpreted in terms of filaments of unstable auroral return current carried by cold ionospheric electrons, and that these cavities are in fact the source elements of sub-structures known to exist within VLF saucers reported previously \citep[e.g.][]{ergun2001,ergun2003}. All of this information suggests that the BBELF waves are more likely to be generated by cold upgoing electron beams rather than by the ion population. The conditions under which BBELF waves initiate the ion heating process, on the other hand, is still an open question, and might depend on many factors, such as thresholds in densities or density gradients, wavelengths, or wave PSDs.  

Another new aspect of the e-POP measurements presented in this study is the correlation of ion heating with downflowing core ions (Figure~\ref{excerpt}). Ion heating events with simultaneous downflowing core ions are more likely in the downward field-aligned current region as shown in Table~\ref{table}. Although similar observations of ion heating associated with downflowing ions at low altitudes do exist from a limit number of sounding rocket experiments \citep{kintner1986,garbe1992, moore1996b,collier2015}, as discussed in the Introduction, there is no statistical study present in the literature to provide a clear picture of this phenomenon. It is rather counterintuitive to observe that the majority of the ion heating events are associated with core ion downflows instead of upflows. The heated ions are usually expected to be accelerated upward by the mirror-force. 

\begin{figure*}
\includegraphics[scale=.5,width=16cm]{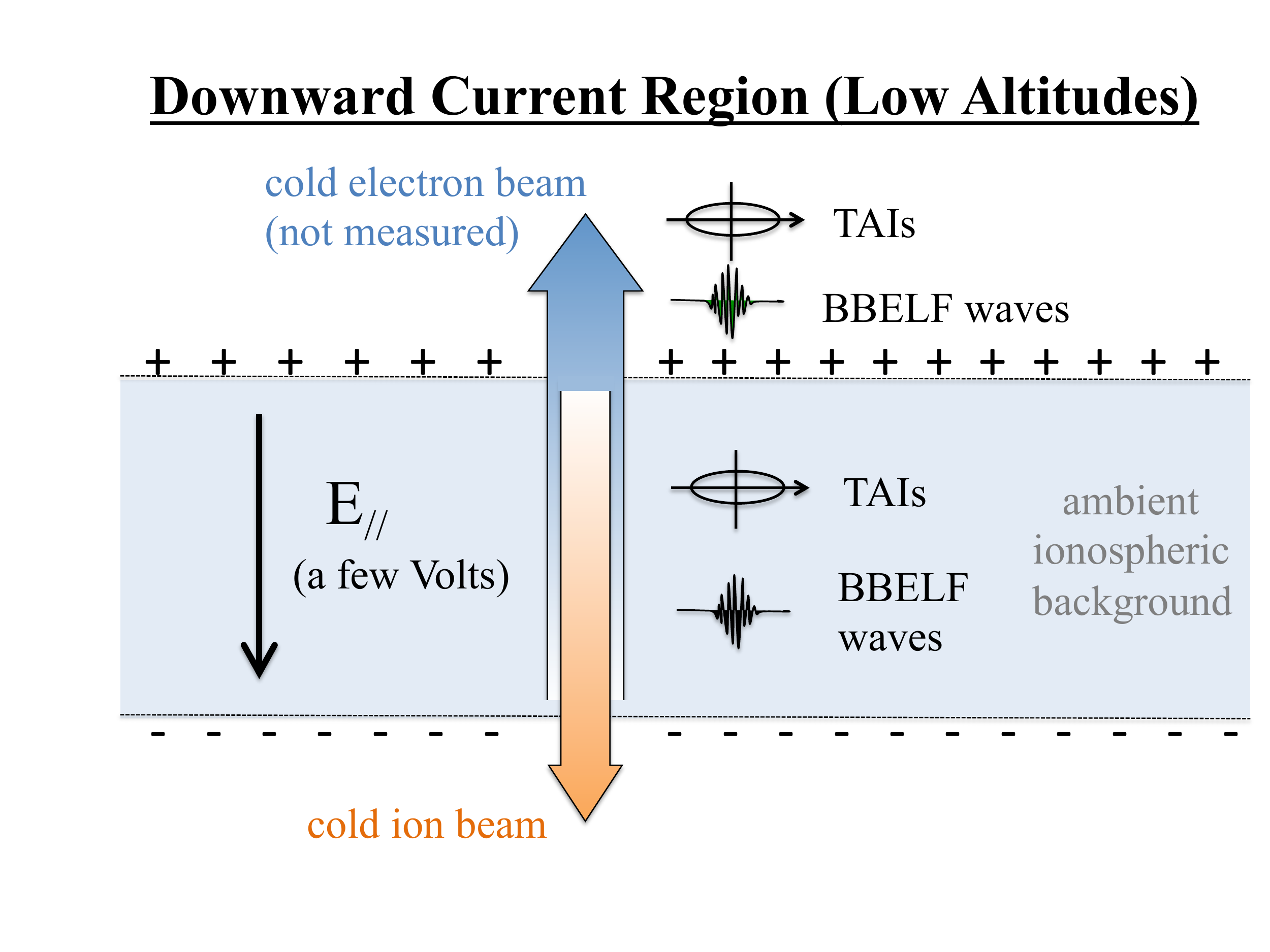}
\caption{Diagram showing proposed physical processes involving O$^{+}$ ion heating, downward core ions, BBELF waves, and downward-pointing electric fields in the return current region as observed by e-POP. The heated core ions are flowing downward due to the downward-pointing electric field. The BBELF waves are generated by instabilities associated with upward cold electron beams. The transversely accelerated ions due to wave-ion heating will move upward once they obtain enough energy to overcome the field-aligned electric potential. This scenario is similar to the so-called ``pressure cooker" model. It is also noted that a density depletion region is expected due to the cold electrons flowing upward and cold ions flowing downward.}
\label{diagram}
\end{figure*}

Although providing a definitive explanation for these downflowing ions within the ion heating region is not the purpose of this study, several possible mechanisms should be mentioned. The most popular one \citep[e.g.][]{loranc1991,endo2000,ogawa2009} is that the downflowing ions are the return of previously upflowing low-energy ion falling back earthward due to gravity. In a dynamic equilibrium, upgoing and downgoing ions are interspersed. However, considering the large convection velocities commonly present in the auroral latitudes, one would expect a large separation between locations of these downflowing ions and the ion heating region that produced them, similar to the picture of the ``cleft ion fountain" \citep{lockwood1985}. Large along-track convection velocities are expected from both the FAI movie (active aurora north-south aligned) and the along-track (perpendicular to $\vec{B}$) first moment calculations (Figure~\ref{irm}a). The one-to-one correlation of ion heating with downflowing ions reported in this study, sometimes in a narrow (near 2 km) region, together with the fact that ion upflows tend to occur just outside the ion heating region, render this mechanism less plausible. A second potential mechanism is that the ion heating takes place in a region which is narrow in altitude. The resulting vertical pressure gradients would generate ion diffusion upward above the heating layer and downward below it. Statistical observations from the EISCAT Svalbard Radar lend some support to this interpretation \citep{buchert2004}. However, ion downflow velocities due to pressure gradients are usually small ($\textless$ 1 km/s) \citep{endo2000,buchert2004} compared to over 2-3 km/s in our study. 

A third candidate, which we favor, is a downward electric field associated with return currents (upward-flowing cold electrons) that occurs low enough in altitude to entrain and accelerate cold, ambient ionospheric ions downward. This is consistent with the fact that the BBELF waves associated with the ion heating are also associated with the return current region \citep{lynch2002}. This is similar to the ``pressure cooker" scenario operating at higher altitudes in the return current region \citep{gorney1985}, where only ions with sufficient energies which can overcome the field-aligned potential barrier are able to flow upward outside the heating region. Figure~\ref{diagram} presents a schematic diagram that summarizes these physical processes in the low-altitude return current region. A density depletion region is expected due to the cold electrons flowing upward and cold ions flowing down.

In the cleft example, we also observe high-energy ions flowing upward just inside the regions in which core ions are flowing downward. In addition, based on wave growth rates calculated from the WHAMP dispersion solver and reported in \citet{knudsen2012}, cold electron beams with energies of the order of 1 eV are sufficient to generate BBELF waves. The corresponding field-aligned potential drop is more than sufficient to accelerate O$^{+}$ ions downward to 2-3 km/s along the same field line. Therefore, the downward-pointing electric fields near the topside ionosphere and above the satellite may explain the correlation of ion heating with core ion downflows instead of upflows. This postulation could be supported by field-aligned electron and electron density measurements, both of which are unavailable for the events reported here.

\section{Conclusion}
Based on observations from e-POP at altitudes between 325 km and 730 km over one year, we have presented a statistical study (24 events) of ion heating and its relation with field-aligned ion bulk flow velocity, low-frequency waves and FACs. Complementary perspectives from several instruments onboard e-POP provide us with new insights into the low-altitude ion energization processes and the field-aligned ion motion. We find that:
\begin{enumerate}
\item  Transverse O$^{+}$ ion heating in the ionosphere and at altitudes as low as 350 km can be intense (up to 4.5 eV), confined to very narrow regions ($\sim$ 2 km across B), is more likely to occur in the downward current region, and is associated with BBELF waves. These BBELF waves are interpreted as linearly polarized in the direction perpendicular to the magnetic field. 
\item The amount of ion heating in these regions cannot be explained by frictional heating, and the correlation of ion heating with BBELF waves suggest that significant wave-ion heating is occurring and even dominating at low altitudes.
\item The minimum altitude of strong wave-ion heating ($\sim$ 350 km) is lower than previously reported in the literature.
\item Contrary to what would be expected from mirror-force acceleration of heated ions, the majority of these heating events (17 out 24) are associated with core ion downflows rather than upflow. This may be explained by a downward-pointing electric field associated with physical processes in the return current region.
\end{enumerate}





\appendix

\section{Sensor potential effect on the FWHM}

\renewcommand\thefigure{A.\arabic{figure}}    
\setcounter{figure}{0}    
\begin{figure}
\includegraphics[scale=.5,width=9cm]{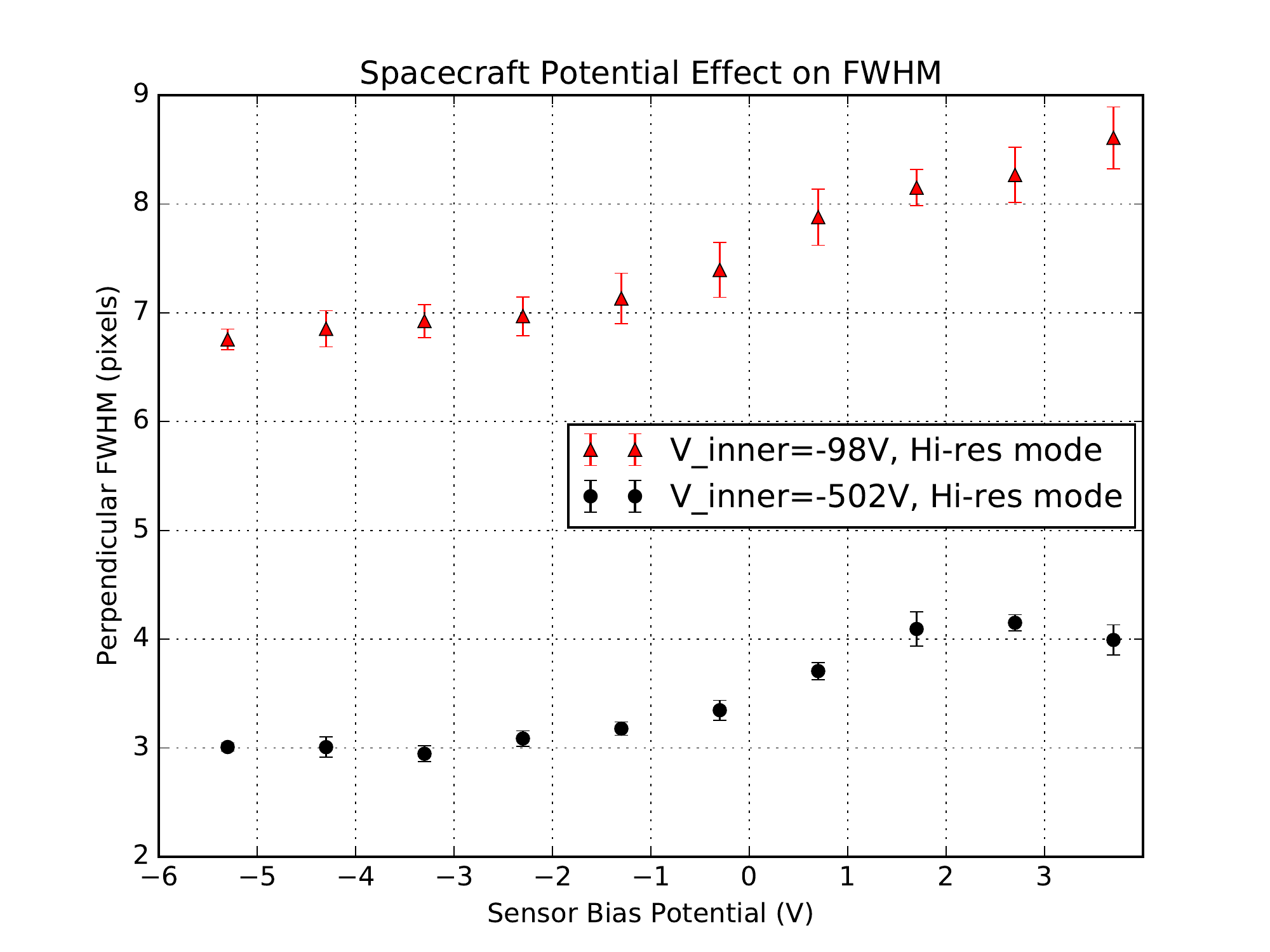}
\caption{Sensor potential effect on the perpendicular FWHM measurements based on a Monte-Carlo simulation. Perpendicular FWHM variations as a function of sensor potential in Hi-res mode are shown when $V_{inner}=-502V$ (black circles) and $V_{inner}=-98V$ (red triangles). }
\label{sensor}
\end{figure}

We estimate the effect of the sensor potential (relative to ambient plasma) on the perpendicular FWHM measurements under different deflection voltages ($V_{inner}$) through forward modeling the distribution using the Monte-Carlo simulation. Figure~\ref{sensor} displays the perpendicular FWHM variations as a function of sensor potential. The perpendicular FWHM profiles are relatively stable when the sensor potentials are negative. A typical value of fixed sensor potential applied onboard is near -5 V. The potential difference between the SEI sensor and the ambient plasma varies mainly due to backscattered and secondary electrons in the nightside auroral region. However, this effect typically results in positive potential change of less than 4 V on the order of the secondary electron temperature in the topside ionosphere where the e-POP satellite resides \citep{whipple1981,garrett1981}. Therefore we conclude that variations in the relative potential between the sensor and the ambient plasma have a negligible effect on the perpendicular FWHM measurements discussed in our study.


%
%
%


\begin{acknowledgments}
This work was supported by a University of Calgary Eyes High Doctoral Recruitment Scholarship and the Natural Sciences and Engineering Research Council of Canada (NSERC). D. M. Miles is supported by faculty startup funding from the University of Iowa. e-POP is funded by the Canadian Space Agency and MDA Corporation. Acknowledgements are owing to GeoForschungsZentrum (GFZ) Potsdam and World Data Center (WDC) for Geomagnetism, Kyoto for providing the Kp index. e-POP data are accessible through http://epop-data.phys.ucalgary.ca/.
\end{acknowledgments}

\end{article}

\end{document}